\documentclass{article}
\usepackage{bm}% bold math
\usepackage{graphicx}
\usepackage[latin1]{inputenc}

\usepackage{apacite}\newcommand{\url}[1]{}

\newcommand{\bd}{\begin{document}}
\newcommand{\ed}{\end{document}}
\newcommand{\beq}{\begin{equation}}
\newcommand{\eeq}{\end{equation}}
\newcommand{\nid}{\noindent}
\newcommand{\ben}{\begin{enumerate}}
\newcommand{\een}{\end{enumerate}}
\newcommand{\bit}{\begin{itemize}}
\newcommand{\eit}{\end{itemize}}
\newcommand{\baR}{\begin{array}}
\newcommand{\eaR}{\end{array}}
\newcommand{\su}{\section}
\newcommand{\ssu}{\subsection}
\newcommand{\sssu}{\subsubsection}
\newcommand{\deq}{\stackrel {\rm def}{=}}
\newcommand{\bea}{\begin{eqnarray}}
\newcommand{\eea}{\end{eqnarray}}
\newcommand{\SOS}[1]{{\bf[#1]}}

\newcommand\D{\displaystyle}

\newcounter{CDef}[section]
\newtheorem{definition}{Definition}%[CDef]
\newcommand{\bdf}{\begin{definition}}
\newcommand{\edf}{\end{definition}}
\newcommand{\bpr} {\noindent {\bf Proof }}
\newcommand{\epr}{\rule{0.25cm}{0.25cm}\\}

\newtheorem{theorem}{Theorem}
\newcommand{\bth}{\begin{theorem}}
\newcommand{\enth}{\end{theorem}}
\newcounter{Cprop}[section]
\newtheorem{proposition}{Proposition}%[Cprop]
\newcommand{\bp}{\begin{proposition}}
\newcommand{\ep}{\end{proposition}}
\newcounter{Ccor}\newtheorem{corrollary}{Corrollary}%[CLem]
\newcommand{\bcor}{\begin{corrollary}}
\newcommand{\ecor}{\end{corrollary}}
\newcounter{CLem}\newtheorem{lemma}{Lemma}%[CLem]
\newcommand{\blem}{\begin{lemma}}
\newcommand{\elem}{\end{lemma}}
\newcounter{Cconj}\newtheorem{conjecture}{Conjecture}%[CLem]
\newcommand{\bconj}{\begin{conjecture}}
\newcommand{\econj}{\end{conjecture}}

\newcommand\bbbr{{\sf I\!R}}

\newcommand\Vr{V_{reset}}
\newcommand\V{{\bf V}}
\newcommand\tV{\tilde{\V}}
\newcommand\Vm{V_{min}}
\newcommand\VM{V_{max}}
\newcommand\hV{\hat{V}}

\newcommand\I{{\bf I}}
\newcommand\J{{\bf J}}
\newcommand\Ia{I^{(ion)}}
\newcommand\Iw{I^{(w)}}
\newcommand\ie{i^{(ext)}}
\newcommand\Ie{{\bf{i}}^{(ext)}}
\newcommand\Is{I^{(syn)}}
\newcommand\Isk{I^{(syn)}_k}
\newcommand\hIa{\hat{I}^{(ion)}}

\newcommand\til{t_i^{(l)}}
\newcommand\tjn{t_j^{(n)}}
\newcommand\tpm{\tau^{\pm}}
\newcommand\bG{\bar{G}}
\newcommand\cA{{\cal A}}
\newcommand\cB{{\cal B}}
\newcommand\cC{{\cal C}}
\newcommand\cD{{\cal D}}
\newcommand\cE{{\cal E}}
\newcommand\cG{{\cal G}}
\newcommand\cH{{\cal H}}
\newcommand\cI{{\cal I}}
\newcommand\cF{{\cal F}}
\newcommand\cN{{\cal N}}
\newcommand\cM{{\cal M}}
\newcommand\cP{{\cal P}}
\newcommand\cS{{\cal S}}
\newcommand\cU{{\cal U}}
\newcommand\cW{{\cal W}}

\newcommand\Bev{\cB(\V,\epsilon)}
\newcommand\bmd{\cB_\cM(\delta)}

\newcommand\ton{\left[\tilde{\omega}\right]_n}
\newcommand\tot{\left[\tilde{\omega}\right]_t}
\newcommand\totm{\left[\tilde{\omega}\right]_{t-1}}
\newcommand\tos{\left[\tilde{\omega}\right]_s}
\newcommand\tom{\tilde{\omega}}
\newcommand\toT{\left[\tilde{\omega}\right]_T}
\newcommand\bom{\mbox{{\boldmath $\omega$}}}
\newcommand\bomt{\bom(t)}
\newcommand\bdom{\bar{\cD}(\bom)}
\newcommand\com{\cC(\bom)}
\newcommand\comt{\cC(\bom(t))}
\newcommand\dom{\cD(\bom)}
\newcommand\mjt{M_j(t,\tom)}

\newcommand\cMo{{\cM_{\bom}}}
\newcommand\cMot{\cM_{\bom(t)}}
\newcommand\Mot{\cM_{\tot}}
\newcommand\Motm{\cM_{\totm}}
\newcommand\MoT{\cM_{\toT}}
\newcommand\Mos{\cM_{\bom(s)}}
\newcommand\oM{\Omega}
\newcommand\SLp{\Sigma_\Lambda^+}

\newcommand\be{{\bf e}}
\newcommand\F{{\bf F}}
\newcommand\FT{\F^T}
\newcommand\FTp{\F^{T+1}}
\newcommand\bcF{\bar{\cF}}
\newcommand\Ft{\F\left[.;.,\tom\right]}
\newcommand\Fto{\F\left[.;t,\tom\right]}
\newcommand\Ftot{\F^{t+1}_{\tom}}
\newcommand\Ftso{\F^{t,s}_{\tom}}
\newcommand\Fkto{F_k\left(t,\tom\right)}
\newcommand\Fkso{F_k\left(s,\tom\right)}
\newcommand\FkVto{F_k\left(t,\tom\right)\V}
\newcommand\FVTo{\F^{t+1}_{\tom}(\V)}
\newcommand\FTo{\F^{t}_{\tom}}
\newcommand\FkTo{F^{t}_{k;\tom}}
\newcommand\DFFTV{\left. D\FTF\right|_{\V}}
\newcommand\DFFTpV{\left. D\FTpF\right|_{\V}}

\newcommand\dAS{d(\oM,\cS)}
\newcommand\dASe{d^{(exp)}(\oM,\cS)}

\newcommand\GFTe{\Gamma_{\cF,T,\epsilon}}
\newcommand\GFTpe{\Gamma_{\cF,{T+1},\epsilon}}
\newcommand\GFTep{\Gamma_{\cF,T,\epsilon'}}

\newcommand\PF{\Pi_{\cF}}
\newcommand\PbF{\Pi_{\bcF}}
\newcommand\VF{{\V_{\cF}}}
\newcommand\VFb{{\V_{\bcF}}}
\newcommand\FF{{\F_{\cF}}}
\newcommand\FTF{{\FT_{\cF}}}
\newcommand\FTpF{{\FTp_{\cF}}}
\newcommand\FFb{{\F_{\bcF}}}
\newcommand\oGF{{\Omega(\GF)}}
\newcommand\oGFTe{{\Omega(\GFTe)}}
\newcommand\oFTe{{\Omega_{\cF}(\GFTe)}}
\newcommand\cPT{\cP^{(T)}}

\newcommand\Jkto{J_k(t,\tot)}

\newcommand\gkt{\gamma_k(t,\tot)}
\newcommand\gks{\gamma_k(s,\tos)}
\newcommand\skt{\sigma_k(t,\tot)}

\newcommand\heff{h^{(eff)}}
\newcommand\PrW{\cP_\cW}
\newcommand\PrI{\cP_{\Ie}}
\newcommand\PrWi{\cP_{\cW,\Ie}}

\newcommand\ltjn{\left\{\tjn\right\}_t}

\newcommand\wsg{\cW^s(\V)}
\newcommand\wsl{\cW^s_{loc}(\V)}
\newcommand\wslf{\cW^s_{loc}(\F(\V))}
\newcommand\wse{\cW^s_{\epsilon}(\V)}
\newcommand\wset{\cW^s_{\epsilon}(\V(t))}

\begin{document}

\title{On Dynamics of  Integrate-and-Fire Neural Networks with Conductance Based Synapses.}

\author{
B. Cessac
\thanks{INRIA, 2004 Route des Lucioles, 06902 Sophia-Antipolis, France.}
\thanks{INLN, 1361, Route des Lucioles, 06560 Valbonne, France.}
\thanks{Universit\'e de Nice, Parc Valrose, 06000 Nice, France.},
T. Vi\'eville
\thanks{INRIA, 2004 Route des Lucioles, 06902 Sophia-Antipolis, France.}
}

\date{\today}

\maketitle

\begin{abstract}

We present a mathematical analysis of a networks with Integrate-and-Fire neurons with conductance based synapses.
Taking into account the realistic fact that the spike time is only known within some \textit{finite} precision,
we propose a model where spikes are effective at times multiple of 
a characteristic time scale $\delta$, where $\delta$ can be \textit{arbitrary} small (in particular, 
well beyond the numerical precision). 
We make a complete mathematical characterization of the model-dynamics
and obtain the following results. The asymptotic dynamics is 
composed by finitely many stable periodic orbits, whose number and  period can be 
arbitrary large and can diverge in a region of the synaptic weights space, traditionally called the ``edge of chaos'', 
a notion mathematically well defined in the present paper. 
Furthermore, except at the edge of chaos, there is a one-to-one correspondence between the membrane potential trajectories and the raster plot.
This shows that the neural code is entirely ``in the spikes'' in this case.
As a key tool, we introduce an order parameter, easy to compute numerically, and closely related to a natural notion of entropy,
 providing a relevant characterization of the computational capabilities of the network.
This allows us to compare the computational capabilities of leaky and Integrate-and-Fire models and conductance based models.
The present study considers networks with constant input, and without time-dependent plasticity, but the framework has been designed for both extensions.	
\end{abstract}

\pagebreak

Neuronal networks have the capacity to treat incoming information,
performing complex computational tasks
(see \cite{rieke-warland:96} for a deep review), including sensory-motor tasks.
It is a crucial challenge to understand how this information
is encoded and transformed. However, when considering \textit{in vivo} neuronal networks,
information treatment proceeds usually from the interaction
of many different functional units having different structures and roles,
and interacting in a complex way. As a result, many time and space scales
are involved. Also, in vivo neuronal systems are not isolated objects and have 
strong interactions with the external world, that hinder
the study of a specific mechanism \cite{fregnac:04}.
In vitro preparations are less subject to these restrictions,
but it is still difficult to design specific neuronal structure
in order to investigate the role of such systems regarding information
treatment \cite{koch-segev:98}. 
In this context models are often proposed,
 sufficiently close from neuronal networks
to keep  essential biological features, but also sufficiently simplified
 to achieve a characterization of their dynamics, the most often  numerically
and, when possible, analytically \cite{gerstner-kistler:02b,dayan-abbott:01}.
This is always a delicate compromise.
 At one extreme, one reproduces all known features of ionic channels, neurons,
synapses ... and lose the hope to have any (mathematics and even numeric) control on what is going on. At the other
extreme, over-simplified models can lose important biological features.
Moreover, sharp simplifications may reveal exotic properties
which are in fact induced by the model itself, but do not
exist in the real system.
This  is a crucial aspect in theoretical
neuroscience, where one must not
forget that models are subject to hypothesis and have
therefore intrinsic limits.

For example, it is widely believed that one of the major advantages of the Integrate-and-Fire (IF) model 
is its conceptual simplicity and analytical tractability that can be used to explore some general principles of neurodynamics 
and coding.  However, though the first IF model was introduced in 1907 by Lapicque \cite{lapicque:07} 
and though many important  analytical and  rigorous results have been published,
 there are essential parts missing in the state of the art in theory
 concerning the dynamics of IF neurons 
(see e.g. \cite{mirollo-strogatz:90,ernst-et-al:95,senn-urbanczik:01,timme-et-al:02,memmesheimer-et-al:06,gong-vanleuwwen:07,jahnke-et-al:08}
and references below for analytically solvable network models of spiking neurons).
 Moreover, while the analysis of an isolated neuron submitted to constant inputs
is straightforward, the action of a periodic current on a neuron reveals already an astonishing complexity and the mathematical analysis requires elaborated methods
from dynamical systems theory \cite{keener-et-al:81,coombes:99,coombes-bressloff:99}. In the same way, the computation of the spike train probability
distribution resulting from the action of a Brownian noise on an IF neuron is not a completely straightforward exercise
\cite{knight:72,gerstner-kistler:02,brunel-sergi:98,brunel-latham:03,touboul-faugeras:07} and may require rather elaborated mathematics.
 At the level of networks the situation is even worse, and the techniques
used for the analysis of a single neuron are not easily extensible
 to the network case. For example, Bressloff and Coombes \cite{bressloff-coombes:00} have 
extended the analysis in \cite{keener-et-al:81,coombes:99,coombes-bressloff:99} to the dynamics of strongly coupled spiking neurons,
but restricted to networks with specific architectures and under restrictive assumptions on the firing times. Chow and Kopell
 \cite{chow-kopell:00}
studied IF neurons coupled with gap junctions but the analysis for large networks assumes constant synaptic weights.
  Brunel and Hakim \cite{brunel-hakim:99} extended the Fokker-Planck analysis combined to a mean-field
approach to the case of a network with inhibitory synaptic couplings but under the assumptions that all synaptic weights are equal.
However, synaptic weight variability plays a crucial role in the dynamics, as revealed e.g. using mean-field methods 
or numerical simulations \cite{vanvresswijk-hansel:97,vanvresswijk-sompolinsky:98,vanvreeswijk:04}. 
 Mean-field methods allow the analysis of networks
with random synaptic weights \cite{amari:72,sompolinsky-et-al:88,cessac-et-al:94,cessac:95,hansel-mato:03,soula-beslon-etal:06,samuelides-cessac:07} 
but they require a ``thermodynamic limit'' where the number of neurons tends to infinity
and finite-size corrections are rather difficult to obtain. Moreover, the rigorous derivation of the mean-field equations,
that requires large-deviations techniques \cite{benarous-guionnet:95}, has not been yet done for the case of IF networks with continuous time dynamics
(for the discrete time case see \cite{soula-beslon-etal:06,samuelides-cessac:07}).   
 
Therefore, the ``analytical tractability'' of IF models is far from being evident.
In the same way, the ``conceptual simplicity''  hides real difficulties which are mainly due to
the following reasons. IF models introduce a discontinuity in the dynamics whenever a neuron crosses
a threshold: this discontinuity, that mimics a ''spike'', maps \textit{instantaneously} the membrane potential
from the  threshold value to a reset value. The conjunction of continuous time dynamics and instantaneous reset
leads to real conceptual and mathematical difficulties. For example, an IF neuron
without refractory period (many authors have considered this situation), can, depending on parameters such as synaptic weights,
 fire uncountably many spikes within a finite time interval, leading
to events which are not measurable (in the sense of probability theory). This prevents the use of standard methods in probability theory
and  notations such as $\rho(t)=\sum_{i=1}^n \delta(t-t_i)$ (spike response function ) simply lose their meaning\footnote{
Obviously, one can immediately point out that (i) this situation is not plausible if one thinks of biological neurons and (ii) is not ``generic''
for IF models. Thus, objection (i) implies that some conclusions drawn from IF models
are not biologically plausible, while objection (ii) needs to be made mathematically clear. This is one of the
goals of this paper.}.  
 Note also that the information theory (e.g. the Shannon
theorem, stating that the sampling period must be less than half the period corresponding to the highest signal frequency) is not applicable with unbounded frequencies.
But IF models have an unbounded frequencies spectrum
(corresponding to instantaneous reset).
 From the information theoretic point of view, it is a temptation to relate this spurious property to the {\em erroneous} fact
that the neuronal network information is not bounded. These few examples illustrate that one should not be abused by the apparent
simplicity of IF models and must be careful in pushing too much their validity in order to explore some general principles of neurodynamics 
and coding.

The situation is not necessarily better when considering numerical implementations of IF neurons. 
Indeed, it is known from a long time that the way the membrane potential is reset in a neuronal network simulation have significant consequences for the dynamics of the model. 
In particular, Hansel et al \cite{hansel-et-al:98} showed that a naive implementation of integrate-and-fire dynamics on a discrete time grid introduces spurious effects and proposed
an heuristic method to reduce the errors induced by time discretization. In parallel, many people have developed
event based integration schemes \cite{brette-rudolph-etal:07},
 using the fact that the time of spike of a neuron receiving instantaneous spikes from other
neurons can be computed analytically, thus reducing consequently the computation time and affording
the simulation of very large networks. 
In addition, 
exact event based computational schemes are typically used for the above-mentioned analytically tractable model classes, see, e.g. 
\cite{mirollo-strogatz:90,timme-et-al:02}.
 Unfortunately, this approach suffers two handicaps. If one considers
more elaborated models than analytically tractable models, one is rapidly faced
to the difficulty of finding an analytical expression for the next spike time \cite{rudolph-destexhe:06}. Moreover,
 any numerical implementations of a neural network model will necessarily introduce errors 
compared to the exact solution. The question is : how does this error
behave when iterating the dynamics ? Is it amplified or damped ? In IF models, as set previously, these errors are due
to the discontinuity in the membrane potential reset and to the time discretization. This has been nicely
discussed by Hansel et al \cite{hansel-et-al:98} paper. These authors point out two important effects. When a neuron
fires a spike between time $t$ and $t+\Delta t$ a local error on the firing time is made when using
time discretization. First, this  leads to an error on the membrane potential and second this error
is propagated to the other neurons via the synaptic interaction term. Unfortunately, this analysis, based on numerical
simulations, was restricted to a specific architecture (identical excitatory neurons) and therefore the conclusions drawn by the
authors
cannot be extended as it is to arbitrary neural architectures. Indeed, as we show in the present paper, the small error induced
by time discretization can be amplified \textit{or damped}, depending on the synaptic weights value.
This leads to the necessity of considering carefully (that is mathematically) the spurious effects
induced by continuous time and instantaneous reset in IF models, as well as the effects of time discretization.
This is one aspect discussed in the present paper. 

More generally, this work contains several conclusions forming a logical chain. 
After a discussion on the characteristic times involved
in real neurons and  comparison to the  assumptions
used in  Integrate and Fire models we argue that discrete time IF models with synchronous dynamics
can be used to model \textit{real neurons} as well, provided that the time scale discretization
is sufficiently small. More precisely, we claim that IF equations are inappropriate if one sticks to
much on the instantaneous reset and spike time, but that they provide a good and mathematically tractable
model if one allows reset and spike to have some duration.  We therefore 
modify the reset and spike definition (while keeping the differential equation  for
the dynamics of the membrane potential below the threshold).
 The goal is however NOT to propose 
yet another numerical scheme for the numerical integration
of continuous time IF models. Instead, our aim
is to analyze mathematically the main properties of the
corresponding dynamical system, describing the evolution
of a \textit{network} with an arbitrary, finite, size (i.e. we don't use neither
a mean-field approach nor a thermodynamic limit). We also consider  an arbitrary architecture.
Finally, in our analysis the time  discretization step is  \textit{arbitrary small}, (thus possibly well below the numerical
precision).
 For this, we use 
a dynamical system approach developed formerly in \cite{blanchard-et-al:00,cessac-et-al:04}.
In particular, in \cite{cessac:07}, a discrete time
version of a leaky Integrate-and-Fire network, was studied. It was shown that  the dynamics is generically
periodic, but the periods can become arbitrary large (in particular, they can be larger
than any accessible computational time) and in (non generic) regions
of the synaptic weights space, dynamics is chaotic. In fact, a complete classification
of the dynamical regimes exhibited by this class of IF models was proposed and a one-to-one correspondence
between membrane potential trajectories and raster plots was exhibited (for recent  contributions
 that study periodic orbits in large networks of IF neurons, see \cite{jahnke-et-al:08,gong-vanleuwwen:07}.).
Beyond these mathematical results, this work warns one about some conclusions drawn
from numerical simulations and emphasizes the necessity to have, when possible,
a rigorous analysis of the dynamics.

The paper \cite{cessac:07} dealt however with a rather simple version of IF neurons (leaky Integrate and Fire)
and one may wonder whether this analysis extend to models closer to biology.
In the present paper  we extend these results, and  give a mathematical treatment of the dynamics of spikes generated in synaptic coupled integrate-and-fire networks
where synaptic currents are modeled in a \textit{biophysically plausible way} (conductance based synapses).
As developed in the text, this extension is far from being straightforward and requires a careful definition
of dynamics \textit{incorporating the integration on the spikes arising in the past}.
This requires a relatively technical construction but this provides 
a setting where a rigorous classification of dynamics arising in IF neural networks with conductance based synapse
can be made, with possible further extended to more elaborated models.\\
  
The paper is organized as follows. In section  \ref{y2modeldef}
we give  a short presentation of
continuous time Integrate-and-Fire models. Then, a careful
discussion about the natural time scales involved
in biological neurons dynamics and how continuous
time IF models violate these conditions is presented.
From this discussion we propose the related discrete time model.
Section \ref{y2dynamics} makes the mathematical analysis
of the model and mathematical results characterizing its dynamics are presented.
 Moreover, we  introduce an order parameter, called $\dAS$, which
measures how close to the threshold
are  neurons during their evolution.
  Dynamics is periodic whenever $\dAS$ is positive,
but the typical orbit period can diverge when it tends
to zero. This parameter is therefore related to an
 effective entropy  within a finite time horizon,
and to the neural network capability
of producing distinct spikes trains. 
In other words, this is a way to measure the ability of the system to emulate different
input-output functions. See \cite{langton:90,bertschinger-natschlager:04} for a 
discussion on the link between the system dynamics and its related computational 
complexity\footnote{It has been proposed that optimal computational capabilities are achieved by
systems whose dynamics is neither chaotic nor ordered but somewhere in-between
order and chaos. This has led to the idea of computation at ``the edge of chaos''.
Early evidence for this hypothesis has been reported by Kauffman \cite{kauffman:69} 
and Langton \cite{langton:90}
considering cellular automata behavior, and Packard \cite{packard:88} using a genetic algorithm.
See \cite{bertschinger-natschlager:04}
for a review. In relation, with these works, theoretical results by Derrida and co-authors 
\cite{derrida-pomeau:86,derrida-flyvbjerg:86} allow to characterize analytically the dynamics
of random Boolean networks and for networks of threshold elements \cite{derrida:87}. Recently
\cite{bertschinger-natschlager:04} have contributed to this question,
considering numerical experiments in the context of real-time computation with recurrent neural networks.}.
The smaller $\dAS$, the larger is the set
of distinct spikes trains that the neural
network is able to produce. This implies in particular
a larger variability in the responses to stimuli.
The vanishing of $\dAS$ corresponds to a region in
the parameters space, called ``the edge of chaos'',
and defined here in mathematically precise way.
In section \ref{y2num} we perform numerical investigations of $\dAS$
in different models from leaky Integrate-and-Fire
to conductance based models. These simulations suggest
that there is a wide region of synaptic conductances
where conductance based models display a large effective
entropy, while this region is thinner for leaky Integrate-and-Fire models. 
This provides a quantitative way to
measuring how conductances based synapses and currents
 enhances the information capacity of integrate and fire models.
Section \ref{y2disc} proposes a discussion
on these results.

\su{General framework.}\label{y2modeldef}

\ssu{General structure of Integrate and Fire models.}\label{GIF}

We  consider the (deterministic) evolution of a set of $N$ neurons.
Call $V_k(t)$ the membrane potential of neuron $k \in \left\{1   \dots N \right\}$ at time
$t$ and let  $\V(t)$ be the vector $\left[V_k(t)\right]_{k=1}^N$. 
We denote by $\V \equiv \V(0)$
the initial condition and the (forward) trajectory of $\V$ by: 
$$\tV \deq \left\{\V(t)\right\}_{t=0}^{+\infty},$$
\nid where time can be either continuous or discrete.
In the examples considered here the membrane potential of all neurons
is uniformly bounded, from above and below, by some values $\Vm,\VM$. Call $\cM=[\Vm,\VM]^N$.
This is the phase space of our dynamical system.

We are focusing here on ``Integrate and Fire models'',
which always incorporate two regimes.
For the clarity of the subsequent developments we briefly review these regimes
(in a reverse order).

\ben
\item \textbf{The ``fire'' regime.}

Fix a real number $\theta \in [\Vm,\VM]$ called the firing threshold of the neuron\footnote{We  assume that all neurons have the
same firing threshold.
The notion of threshold is already an approximation which 
is not sharply defined in Hodgkin-Huxley \cite{hodgkin-huxley:52} or Fitzhugh-Nagumo \cite{fitzHugh:61,nagumo-et-al:62} models (more precisely
it is not a constant but it depends on the dynamical variables). 
Recent experiments \cite{naundorf-et-al:06,cormick-et-al:07,naundorf-et-al:07} even suggest
that there may be no real potential threshold.}.
Define the firing times of neuron $k$, for the trajectory\footnote{Note that, since the dynamics
is deterministic, it
is equivalent to fix the forward trajectory  or the initial condition $\V\equiv \V(0)$. }  $\V$, by:

\beq\label{Tdech}
t_k^{(n)}(\V)=\inf\left\{t  \ | t > t_k^{(n-1)}(\V), \ V_k(t) \geq \theta  \right\}
\eeq

\nid where $t_k^{(0)}=-\infty$.
The firing of neuron $k$ corresponds to  the following procedure. 
If $V_k(t) \geq \theta$ then neuron membrane potential is reset 
\textit{instantaneously} to some \textit{constant} reset value $\Vr$ and a spike is emitted 
toward post-synaptic neurons.
In mathematical terms firing reads\footnote{Note that the firing condition includes the possibility to have
a membrane potential value above the threshold. This extension
of the standard definition affords some discontinuous jumps in the dynamics.
These jumps arise when considering addition of (discontinuous) noise, or 
$\alpha$ profiles with jumps (e.g. $\alpha(t) = \frac{1}{\tau}e^{-\frac{t}{\tau}}, \ t \geq 0$).
They also appear when considering a discrete time evolution.
Note that strictly speaking, this can happen, within the numerical precision, even with  numerical
schemes using interpolations to locate more precisely the spike time \cite{hansel-et-al:98}. \label{Notejump} }:

\beq\label{IFt}
V_k(t) \geq \theta \Rightarrow V_k(t^+) = \Vr
\eeq

\nid where $\Vr \in [\Vm,\VM]$ is called the ``reset potential''.
 In the sequel we  assume, without loss of generality,
 that $\Vr=0$. This reset has a dramatic effect.
Changing the initial values of the membrane potential, one may expect
some variability in the evolution. Now, fix a neuron $k$  and  assume that 
there is a time $t >0$ and an interval $[a,b]$ such that, $\forall V_k(0) \in
[a,b]$,  $V_k(t) \geq \theta$. 
 Then, after reset,
this interval is mapped to  the point $\Vr$.
Then, all trajectories born from $[a,b]$
collapse on the same  point and have obviously the same further
evolution. 
Moreover, after reset, the membrane potential evolution does not depend
on its past value. This induces an interesting property used in all the Integrate
and Fire models that we know (see e.g. \cite{gerstner-kistler:02b}).  
\textit{The dynamical evolution is essentially determined
by the firing times of the neurons, instead of their membrane potential
value.}

\item\textbf{ The ``Integrate regime''.} 

Below the threshold, $V_k<\theta$, neuron $k$'s dynamics 
is driven by an equation of form: 

\beq \label{EqQCV}
C\frac{dV_k}{dt}+g_k V_k=i_k,
\eeq

\nid where $C$ is the membrane capacity of neuron $k$. Without loss of generality we normalize the quantities and fix $C=1$.
In its most general form, the neuron $k$'s membrane conductance $g_k > 0$ depends on $V_k$ (see e.g. Hodgkin-Huxley equations \cite{hodgkin-huxley:52})
and time $t$, while the current $i_k$ can also depend on $\V$, the membrane potential \textit{vector}, on time $t$, and
also on the collection of past firing times.
The current $i_k$  can include various phenomenological terms.
Note  that (\ref{EqQCV}) deals with neurons considered as points instead of
spatially extended objects. 

\een

Let us give two examples investigated in this paper.

\ben 

\item[(i)] \textbf{The leaky integrate and fire model.}

In its simplest form equation (\ref{EqQCV})
reads:

\beq\label{LIF}
\frac{dV_k}{dt}=-\frac{V_k}{\tau_k}+i_k(t)
\eeq

\nid where $g_k$ is a constant, and $\tau_k=\frac{g_k}{C}$
is the characteristic time for membrane potential
decay when no current is present.
This model has been introduced in \cite{lapicque:07}.

\item[(ii)] \textbf{Conductance based models with $\alpha$ profiles} 

More generally, conductance and currents depend on $\V$ only via the
previous \textit{firing times} of the neurons \cite{rudolph-destexhe:06}.
Namely, conductances (and currents) have the general form\footnote{The rather cumbersome notation $g_k(t,\ltjn)$
simply expresses that in conductance based models the conductance depends
on the \textit{whole set (history)} of (past) firing times. Note that membrane potentials are reset after neuron firing, 
but \textit{not} neuron conductances.},
  $g_k \equiv g_k(t,\ltjn)$ where $\tjn$ is the $n$-th firing time of
neuron $j$ and $\ltjn$ is the list of firing times of all neurons up
to time $t$. This corresponds to the fact that the occurrence of a post-synaptic potential on synapse $j$, at time $\tjn$,
results in a change of the conductance $g_k$ of neuron $k$. As an example, we consider models of form:

\beq\label{yvettenet}
\frac{dV_k}{dt}=-\frac{1}{\tau_L} \, (V_k -E_L)-i_k^{(syn)}(V_k,t,\ltjn)+i_k^{(ext)}(t)
\eeq

\nid where the first term in the r.h.s. is a leak term, and where the synaptic
current reads:

$$i^{(syn)}(V_k,t,\ltjn)=(V_k-E^+)\, \sum_{j=1}^N g_{kj}^+(t,\ltjn) + (V_k-E^-)\, \sum_{j=1}^N g_{kj}^-(t,\ltjn),$$

\nid where $E^\pm$ are reversal potential (typically $E^+ \simeq 0 mV$ and $E^- \simeq -75 mV$)
and where:

$$g_{kj}^{\pm}(t,\ltjn)=  G_{kj}^{\pm} \sum_{n=1}^{M_j(t,\V)} \alpha^{\pm}(t-\tjn).$$ 

In this equation, $M_j(t,\V)$ is the number\footnote{Henceforth, one assumes that there are finitely
many spikes within a finite time interval. For continuous time dynamics, this fact is not guaranteed when neglecting
the refractory period. Note also that this number, as well as the list $\ltjn$, depends on the initial condition $\V$
and a small change in the initial condition may induce a drastic change of $M_j(t,\V)$ at time $t$, as discussed later. This effect
is sometimes disregarded \cite{coombes:99}.  This issue has also
been discussed (for current based IF-like models) as "phase history functions" in \cite{ashwin-timme:05,broer-et-al:08} (we thank one
of the reviewers for this remark).}
of times neuron $j$ has fired at time $t$.
$G_{kj}^{\pm}$ is the synaptic efficiency (or synaptic weight) of
the synapse $j \to k$. (It is zero if there is no synapse $j \to k$),
 where $+$ [$-$] expresses that synapse $j \to k$ is excitatory [inhibitory]. 
The $\alpha$ function mimics the conductance time-course
after the arrival of a post-synaptic potential. A possible choice is:

\beq\label{alpha}
\alpha^\pm(t)= H(t) \, \frac{t}{\tpm} \, e^{-\frac{t}{\tpm}},
\eeq

\nid with $H$ the Heaviside function and $\tpm$ being characteristic times.
This synaptic profile, with $\alpha(0)=0$ while $\alpha(t)$ is maximal
for $t=\tau$, allows us to smoothly delay the spike action on the post-synaptic
neuron. We are going to neglect other forms of delays in the sequel.

Then, we may write (\ref{yvettenet}) in the form (\ref{EqQCV}) with:

\beq\label{gklt}
g_k(t,\ltjn)=\frac{1}{\tau_L}
+\sum_{j=1}^N  G_{kj}^{+} \sum_{n=1}^{M_j(t,\V)} \alpha^{+}(t-\tjn)
+\sum_{j=1}^N  G_{kj}^{-} \sum_{n=1}^{M_j(t,\V)} \alpha^{-}(t-\tjn),
\eeq

\nid and:

\beq\label{ik}
i_k(t,\ltjn)=\frac{E_L}{\tau_L} 
+  E^{+} \, \sum_{j=1}^{N} g_{kj}^{+}(t,\ltjn)
+  E^{-} \, \sum_{j=1}^{N} g_{kj}^{-}(t,\ltjn)
+i^{(ext)}_k(t).
\eeq

\een

\ssu{Discrete time dynamics.}

\sssu{Characteristic time scales in neurons dynamics.}\label{StimeScales}

Integrate and Fire models assume an instantaneous reset of the membrane potential
corollary to an infinite precision for the spike time. We would like to discuss
shortly this aspect.
Looking at the spike shape reveals some natural time scales:
the spike duration $\tau$  (a few ms);
the refractory period $r \simeq 1 ms$; and the spike time precision. 
Indeed, one can mathematically define
the spike time as the time where the action potential reaches
some value (a threshold, or the maximum of the membrane potential during the spike), but, on practical ground,
spike time is not determined with an infinite precision.
An immediate conclusion is that it is not correct, from an operational point of
view, to speak about the ``spike time'', unless one precises that 
this time is known with a finite precision $\delta \tau$. 
Thus the notion of list of firing time $\ltjn$ used in section \ref{GIF},
must be revisited, and a related question is ``what is the effect of this indeterminacy on the dynamical evolution ?''.
Note that this (evident ?) fact  is
forgotten when modeling e.g. spike with Dirac distributions.
This is harmless as soon as the characteristic time $\delta \tau$ is smaller than all other characteristic
times involved in the neural network. This is essentially true in biological
networks but \textit{it is not true} in Integrate and Fire models.

These time scales arise when considering \textit{experimental} data on spikes.
When dealing with \textit{models}, where membrane potential dynamics is represented
by  ordinary differential equations usually derived from Hodgkin-Huxley model, other implicit
times scales must be considered. Indeed, Hodgkin-Huxley formulation
in term of ionic channel activity assumes  an integration over a time
scale $dt$ which has to be (i) quite larger than the characteristic time scale $\tau_P$ of opening/closing 
of the channels, ensuring that the notion of \textit{probability} as
a meaning; (ii) quite larger than the correlation time $\tau_C$ between 
channel states ensuring that the \textit{Markov} approximation
used in the equations of the variable $m,n,h$ is legal.
This means
that, although the mathematical definition of $\frac{d}{dt}$ assumes a limit $dt \to 0$,
there is a time scale below which the ordinary differential equations lose
their meaning. Actually, the mere notion of ``membrane potential'' already assumes
an average over  microscopic time and space scales.
Note that the same is true for all differential 
equations in physics ! But this (evident ?) fact is sometimes
forgotten when dealing with Integrate and Fire models. Indeed, to summarize, the range
of validity of an ODE modeling membrane potential dynamics 
is $\max(\tau_C,\tau_P) << dt << \delta \tau << \tau$.
But the notion of \textit{instantaneous reset}
implies  $\tau=0$ and the mere notion of \textit{spike time} implies that $\delta \tau=0$ !!

There is a last time scale related to the notion of \textit{raster
plot}. It is widely admitted that the ``neural code'' is contained
in the spike trains. Spike trains are represented by raster plots,
namely  bi-dimensional diagrams with time on abscissa and
some neurons labeling on ordinate.
If neuron $k$ fires a spike ``at time $t_k$''
 one represents a vertical bar at the point ($t_k,k$).
Beyond the discussion above on the spike time precision,
the physical
measurement of a raster plot involves a time discretization 
corresponding
to the time resolution $\delta_A$ of the apparatus. When observing a set of neurons
activity, this introduces an apparent synchronization,
since neurons firing between $t$ and $t+\delta_A$ will be considered
as firing simultaneously. This raises several deep questions. In
such circumstances the ``information'' contained in the observed raster plot
depends on the time resolution $\delta_A$ \cite{golomb-et-al:97,panzeri-treves:96} and it should increase as $\delta_A$
decreases. But is there a limit time resolution below which this information
does not grow anymore ?
  In Integrate and Fire models this limit
is $\delta_A=0$. This may lead to the conclusion
that neural networks have an unbounded information capacity.
But is this a property of \textit{real} neurons or only of Integrate and Fire \textit{models} ?\\

The observation of raster plots corresponds to switching 
from the continuous time dynamics of membrane potential
to the discrete time and synchronous dynamics of spike trains.
One obtains then, in some sense, a new dynamical system,
of symbolic type, where variables are bits ('0' for no spike,
and '1' otherwise). The main advantage of this new dynamical
system is that it focuses on the relevant variables
as far as information and neural coding is concerned i.e.
one focuses on spikes dynamics instead of membrane potentials.
In particular, membrane potentials may still depend continuously
on time, but one is only interested in their values at the times
corresponding to the time grid imposed by the raster plot measurement.
In some sense this produces a stroboscopic dynamical system,
with a frequency given by the time resolution $\delta_A$,
producing a phenomenological representation of the underlying
continuous time evolution.  

This has several advantages.
(i) this simplifies the mathematical analysis of the dynamics
avoiding the use of delta distributions, left-right limits,
etc ... appearing in the continuous version;
(ii) provided that mathematical results do not
depend on the finite time discretization scale,
one can take it arbitrary small;
(iii) it enhances the role of symbolic coding and raster plots.

Henceforth, from now on, we fix a positive
time scale $\delta>0$ which can be mathematically arbitrary small,
such that (i) a neuron can fire at most once between
$[t, t+\delta[$ (i.e. $\delta << r$, the refractory period);
(ii) $dt << \delta$, so that we can keep the continuous time evolution of membrane 
potentials (\ref{EqQCV}), taking into account
time scales smaller than $\delta$, and 
 integrating membrane potential dynamics on the intervals $[t,t+\delta[$;
(iii) the spike time is known within a precision $\delta$. Therefore,
the terminology, ``neuron $k$ fires at time $t$'' has to be replaced
by ``neuron $k$ fires between $t$ and $t+\delta$''; (iv)
the update of conductances is made at times multiples\footnote{This
could correspond to the following ``experiment''. Assume
that we measure the spikes emitted by a set of in vitro neurons,
and that we use this information to update the conductances of
a model like (\ref{yvettenet}), in order to see how this model
``matches'' the real neurons (see  \cite{jolivet:06} for a nice
investigation in this spirit). Then, we would have to take into account
that the information provided by the experimental raster plot is discrete, with a clock-based grid,
even if the membrane potential evolves continuously.} of $\delta$. 

\sssu{Raster plot.} \label{Rast}

In this context, we introduce a notion of ``raster plot'' which is essentially
the same as in biological measurements.
A raster plot is a sequence $\tom \deq \left\{\bom(t)\right\}_{t=0}^{+\infty}$,
of vectors $\bom(t) \deq \left[\omega_k(t) \right]_{k=1}^N$
  such that the entry $\omega_k(t)$
is $1$ if neuron $k$ fires between $[t,t+\delta[$ and is $0$ otherwise.
Note however that for mathematical
reasons, explained later on, a raster plot corresponds to the list of firing states $\left\{\bom(t) \right\}_{t=0}^\infty$ 
over an  \textit{infinite time horizon}, while on practical grounds one always considers bounded times. 

Now, for each $k=1 \dots N$, one can decompose the interval $\cI = [\Vm,\VM]$ into
$\cI_0 \cup \cI_1$ with $\cI_0=[\Vm,\ \theta[$, $\cI_1=[\theta,\VM]$. If
 $V_k \in \cI_0$ neuron $k$ is \textit{quiescent}, otherwise it \textit{fires}. 
This splitting induces a partition $\cP$ of $\cM$, that we call the ``natural partition''.
The elements of $\cP$ have the following form. Call $\Lambda=\left\{0,1 \right\}^N$.
Let $\bom=\left[\omega_k\right]_{k=1}^N \in \Lambda$. This is a $N$ dimensional vector 
with binary components $0,1$. We call such a vector a \textit{firing state}.
Then $\cM = \D{\bigcup_{\bom \in  \Lambda} \cMo}$ where:

\beq\label{PartM}
\cMo = \left\{\V \in \cM \ | \ V_k \in \cI_{\omega_k}  \right\}.
\eeq

Therefore the partition $\cP$ corresponds to classifying the membrane potential vectors
according to their firing state. Indeed, to each point $\V(t)$ of the trajectory $\tV$ corresponds
a firing state $\bom(t)$ whose components are given by:

\beq\label{Defomi}
\omega_k(t)=Z[V_k(t)],
\eeq 

\nid where $Z$ is defined by :

\beq\label{Z}
Z(x)=\chi\left[x \geq \theta \right],
\eeq

\nid where $\chi$ is the indicator function that  will later on allows us to include the firing condition in the evolution equation
of the membrane potential (see (\ref{DNN})).
  On a more fundamental ground, the introduction of raster plots leads
to a switch from  the dynamical description of neurons,
 in terms of their membrane potential evolution, to a description in terms of \textit{spike trains} 
where $\tom$ provides a natural ``neural code''.
 From the dynamical systems point of view, it introduces formally a symbolic
coding and symbolic sequences are easier to handle than 
continuous variables, in many aspects such as the computation of
topological or  measure theoretic quantities like topological or Kolmogorov-Sinai entropy \cite{Katok-Hasselblatt:98}.
 A natural related question is whether there is a
one-to-one correspondence between the membrane potential trajectory
and the raster plot (see theorem \ref{ThdAS}).

Note that in the deterministic models that we consider here, the evolution, including 
the firing times of the neurons and the raster plot, is entirely determined by the initial conditions.
Therefore, there is no need to introduce an exogenous process (e.g. stochastic) 
for the generation of spikes (see \cite{destexhe-contreras:06} for a nice discussion on these aspects). \\

Furthermore, this definition has a fundamental consequence
In the present
context, current and conductances at time $t$ become
functions of the raster plot up to time $t$. Indeed, 
we may write (\ref{gklt}) in the form:

\beq\label{gktomint}
 g_k(t,\tot) \deq \sum_{j} \bG_{kj}^\pm \sum_{n=1}^{\mjt} \alpha^\pm(s-\tjn) ds
\eeq

\nid where $\tot = \left\{\bom(s)\right\}_{s=0}^{t}$ is the raster plot up to time $t$
and $\mjt$ is the number of spikes emitted by neuron $j$ up to time $t$, in the raster plot $\tom$ (i.e.
$\mjt=\sum_{s=1}^t \omega_j(s)$).
But now $\tjn$ is  a multiple of $\delta$. \\

\textbf{Remark.} In continuous time Integrate and Fire
models $\tom$  can assume uncountably many values. This is because
a neuron can fire at any time and because firing is instantaneous. Therefore,
the  same property holds also
 if one considers sequences of firing states over a \textit{bounded time horizon}. 
This is still the case even if one introduces a refractory 
period, because even if spikes produced by a given neuron are separated by a time
slot larger or equal than the refractory period, the first spike can occur at \textit{any
time} (with an infinite precision).  If, on the opposite, we discretize time
with a time scale $\delta$ small enough to ensure that each neuron can
fire only once between $t$ and $t+\delta$, $\tom$,
truncated at time $T\delta$ can take at most $2^{NT}$ values. 
For these reasons, the ``computational power'' of 
Integrate and Fire models with continuous time is sometimes considered as infinitely larger 
than a system with clocked based discretization \cite{maass-bishop:03}.
The question is however whether this computational power is
something that real neurons have, or if we are dealing with
a model-induced property.

\sssu{Integrate regime.}\label{SDIF}

For this regime, as we already mentioned, we keep the possibility to have a continuous time ($dt << \delta$)
evolution of membrane potential (\ref{EqQCV}). This allows us
to integrate $V$ on time scales smaller than $\delta$. But, since
conductances and currents depends now on the raster plot $\tom$, we may now write (\ref{EqQCV})
in the form:

\beq\label{IFu}
\frac{dV_k}{dt}+ g_k(t,\tot) \, V_k= i_k(t,\tot); \qquad \mbox{whenever} \quad V_k < \theta.
\eeq

When neuron $k$ does not fire 
between $t,t+\delta$ one has, integrating the membrane potential on
the interval $t,t+\delta$ (see appendix):

\beq 
V_k(t+\delta)= \gamma_k(t,\tot) \, V_k(t)+\Jkto,
\eeq

\nid where:

\beq \label{gamma}
\gamma_k(t,\tot) \deq e^{-\int_{t}^{t+\delta} \, g_k(s,\tot) \, ds},
\eeq

\nid and where:

\beq\label{Jkt}
\Jkto=\int_{t}^{t+\delta} i_k(s,\tot) \, \nu_k(s,t+\delta,\tot) \, ds,
\eeq

\nid is the integrated current with:

\beq\label{nuND}
\nu_k(s,t+\delta,\tot)=e^{-\int_{s}^{t+\delta} \, g_k(s',\tot) \, ds'}.
\eeq

\textbf{Remarks.} 

\ben

\item In the sequel, we  assume that the \textit{external} current (see (\ref{ik})) is time-\textit{constant}.
Further developments with a time dependent current, i.e. in the framework of an input-output computation \cite{bertschinger-natschlager:04}, will be considered next.

\item We note the following property, central in the subsequent developments.
Since $g_k(t,\tot)>0$,

\beq\label{gammacont}
\gamma_k(t,\tot) < 1, \forall t, \  \forall \tom, \  \forall k.
\eeq
\een

\sssu{Firing regime.}

\begin{figure}[ht!]
\begin{center}

\includegraphics[height=3cm,width=3cm,clip=false]{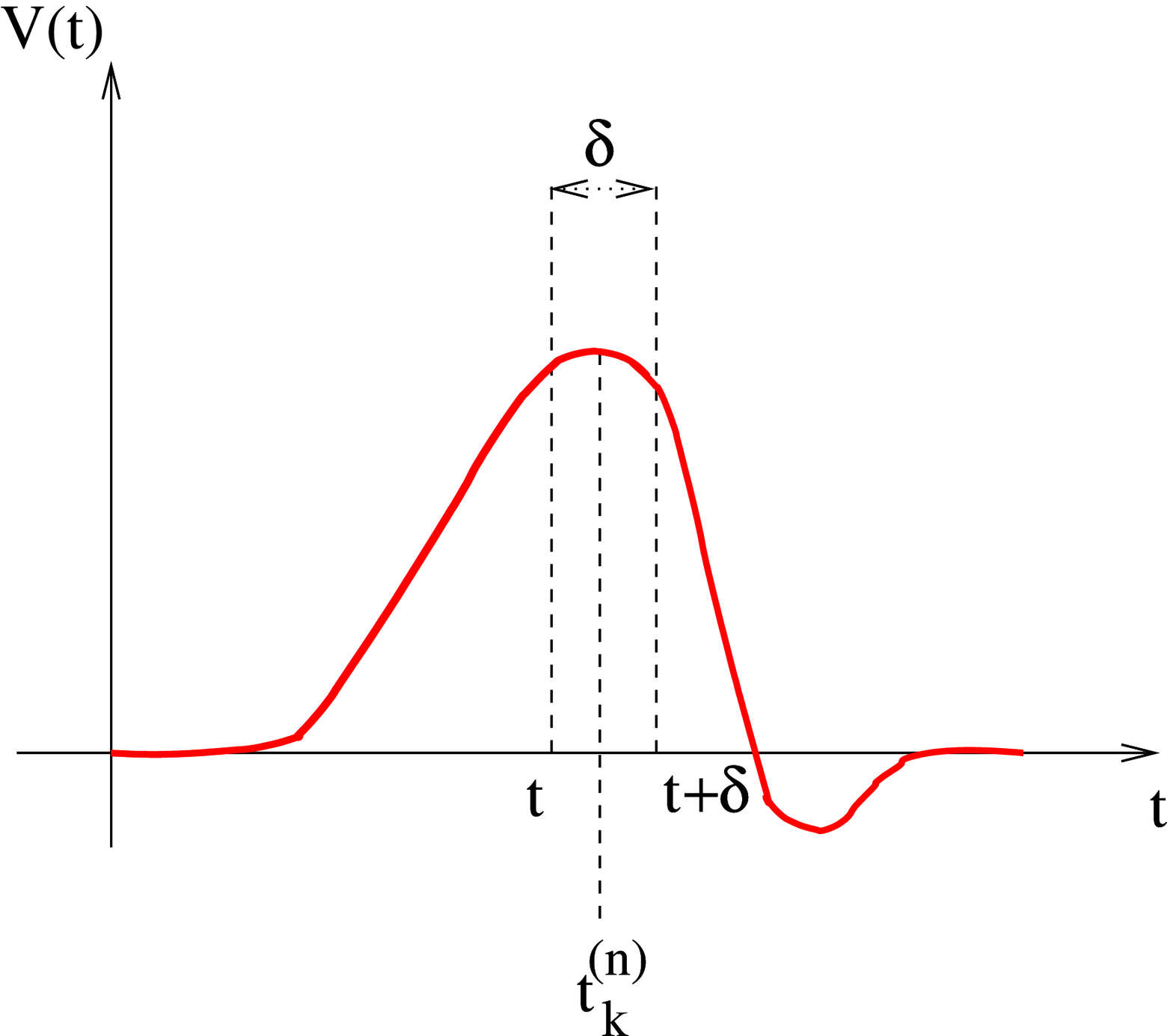}
\hspace{0.3cm}
\includegraphics[height=3cm,width=3cm,clip=false]{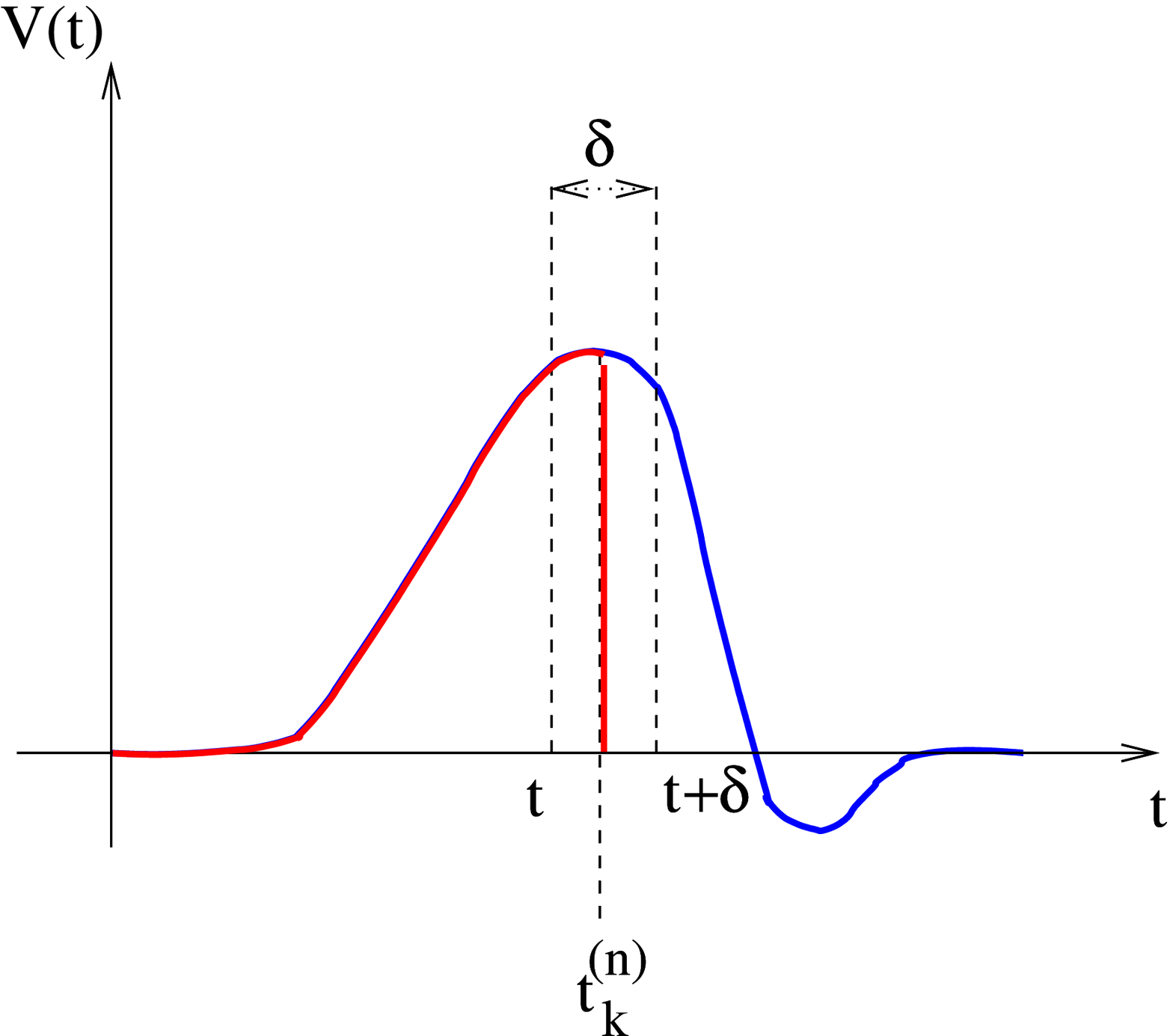}
\hspace{0.3cm}
\includegraphics[height=3cm,width=3cm,clip=false]{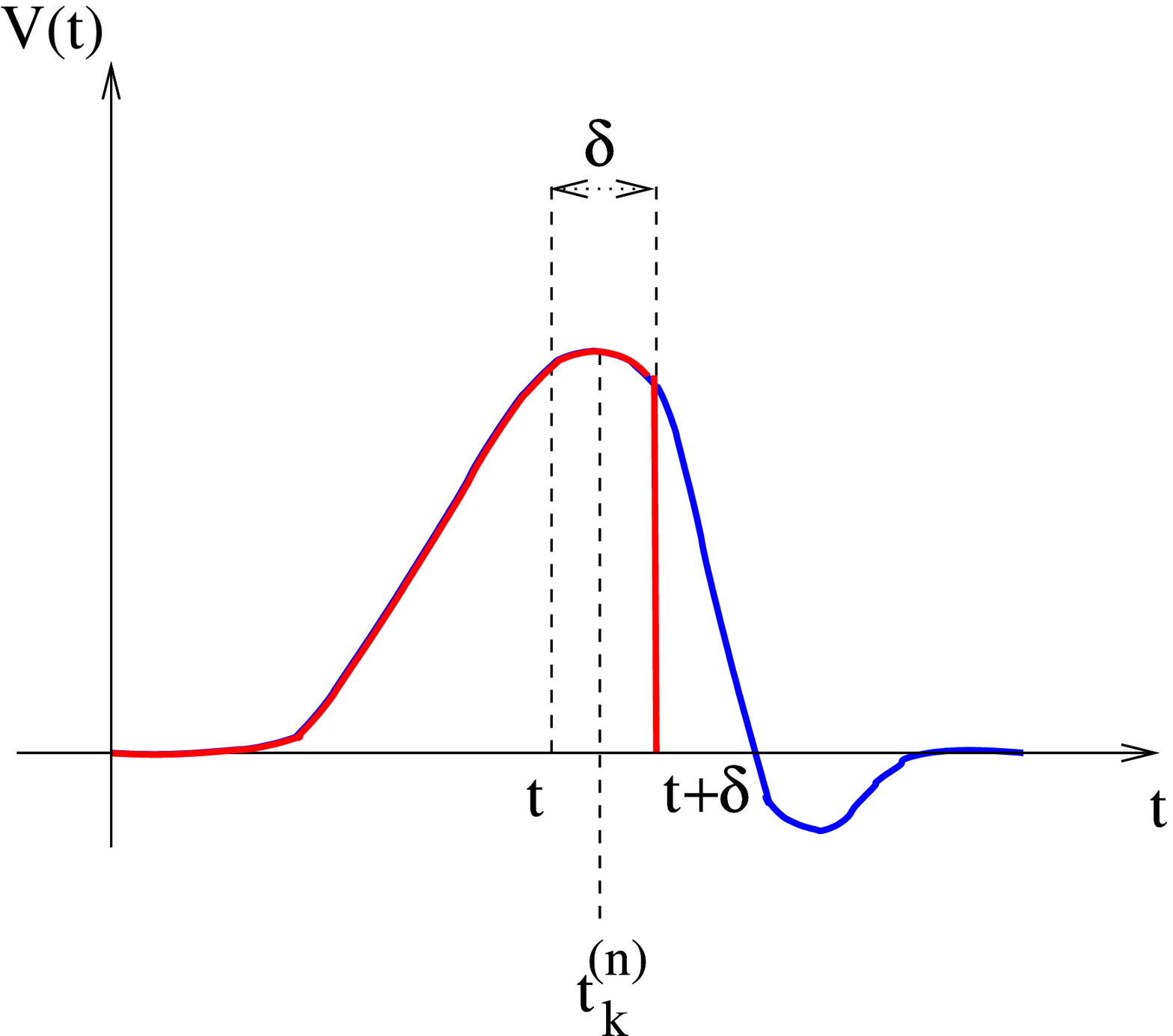}
\hspace{0.3cm}
\includegraphics[height=3cm,width=3cm,clip=false]{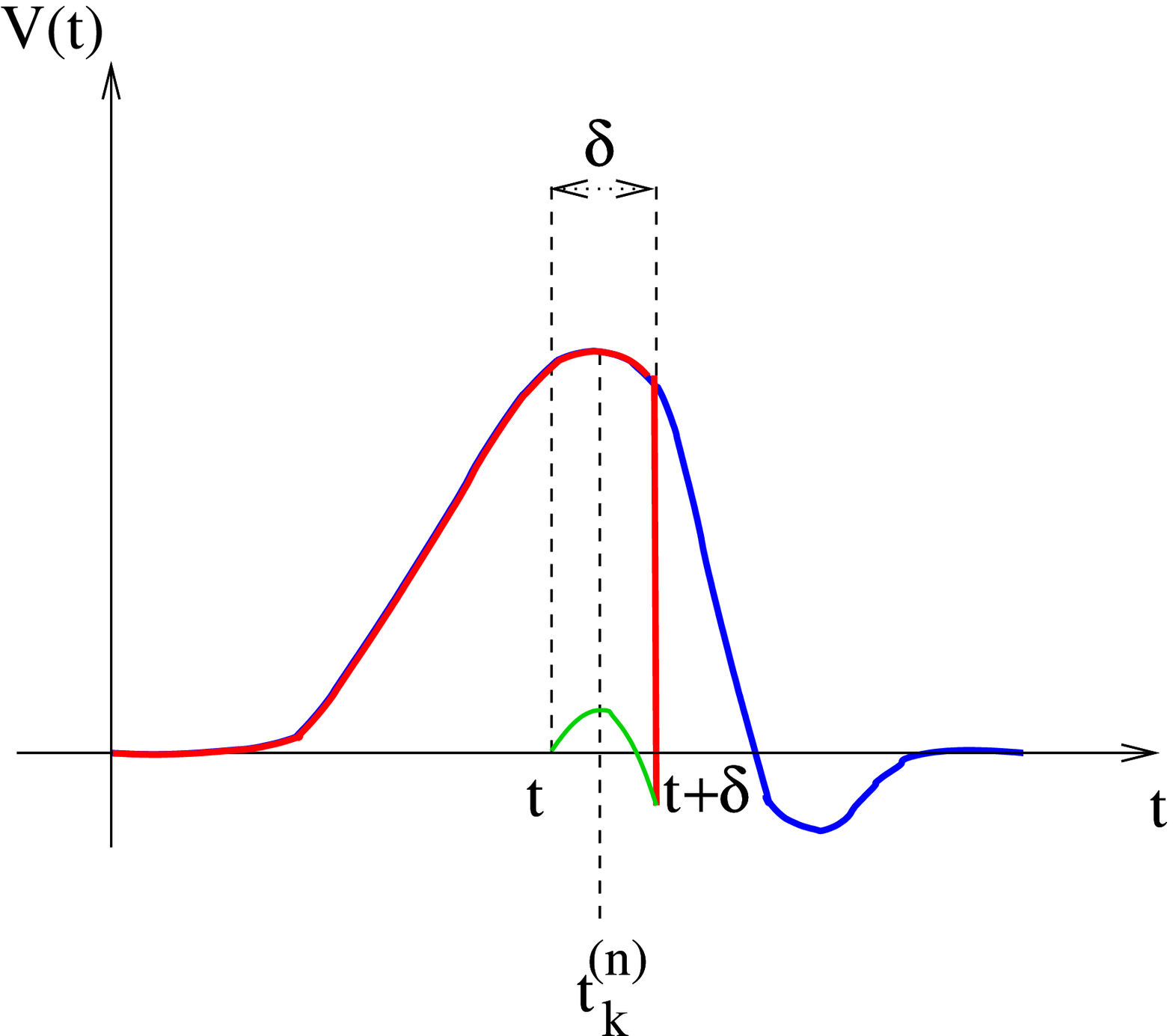}

\vspace{0.5cm}
\caption{A. ``Real spike'' shape; the sampling window is represented at a scale corresponding to a ``small'' sampling rate to enhance the related bias. 
         B. Spike shape for an integrate and fire model with instantaneous reset, the real shape is in blue. 
         C. Spike shape when reset occurs at time $t+\delta$ (alternative I). 
         D. Spike shape with reset at time $t+\delta$ plus addition of the integrate current (green curve) (alternative II). }
\label{Fspikes}\end{center}
\end{figure}

Let us now consider the case where neuron $k$ fires between $t$ and $t+\delta$.
In classical IF models this means that there is some $t_k^{(n)} \in [t,t+\delta[$ such that $V_k(t_k^{(n)})=\theta$.
Then, one sets $V_k(t_k^{(n)+})= \Vr$ (instantaneous reset). This corresponds
to Fig.~\ref{Fspikes}b. Doing this one makes some error compared to the real
spike shape depicted in Fig.~\ref{Fspikes}a. In our approach, one does not know
exactly when firing occurs but we use the approximation that 
the spike is taken into account at time $t+\delta$. This means
that we integrate $V_k$ until $t+\delta$ then reset it. In this scheme
$V_k$ can be larger than $\theta$ as well. This explains
why $Z(x)=\chi(x \geq\theta)$. This procedures corresponds
to Fig.~\ref{Fspikes}c (alternative I). One can also use a slightly different
procedure. We reset the membrane potential at $t+\delta$
but we add to its value the integrated current between $[t,t+\delta[$.
This corresponds to Fig.~\ref{Fspikes}d (alternative II). We have therefore three types
of approximation for the real spike in Fig.~\ref{Fspikes}a. Another
one was proposed by \cite{hansel-et-al:98}, using a linear interpolation
scheme. Other schemes could be proposed as well. One expects them
to be all equivalent when $\delta \to 0$. For finite $\delta$,
the question whether the error induced by these approximations
is crucial and discussed in section \ref{RqTdisc}. 

In this paper we shall concentrate on alternative II though mathematical
results can be extended to alternative I in a straightforward way.
This  corresponds to the initial  choice of the Beslon-Mazet-Soula Model motivating
the paper \cite{soula-beslon-etal:06} and the present work. 

In this case, the reset corresponds to :

 \beq\label{IFtD}
V_k(t) \geq \theta \Rightarrow V_k(t+\delta) = \Jkto,
\eeq

 \nid (recall that $\Vr=0$).

Integrate and Fire regime can now be included in a unique equation,
using the function $Z$ defined in (\ref{Z}):

\beq \label{DNN}
V_k(t+1)= \gamma_k(t,\tot)\left[1-Z(V_k(t))\right]V_k(t)+\Jkto,
\eeq
 
\nid where we set $\delta=1$ from now on.

%
%%%%%%%%%%%%%%%%%%%% Cas 1cdf
%%%%%%%%%%%%%%%%%%%%%%%%%

\ssu{Examples.}

\sssu{The Beslon-Mazet-Soula Model (BMS).}

Consider the leaky integrate and fire model, where
conductances are constant. Set $W_{kj}=G_{kj}^+E^+$ ($W_{kj}=G_{kj}^-E^-$)
for excitatory (inhibitory) synapses.
Then, replacing the $\alpha$-profile by a Dirac distribution,
 (\ref{DNN}) reduces to:

\beq  \label{DND}
V_k(t+1) = \, \gamma \, V_k(t) \, \left[1-Z(V_k(t))\right]+ \sum_{j=1}^N W_{kj} \, Z(V_j(t)) + \ie_k 
\eeq

 This model
has been proposed by  G. Beslon,  O. Mazet and H. Soula in \cite{soula-beslon-etal:06}.
A mathematical analysis of its asymptotic dynamics has been done 
in \cite{cessac:07} and we  extend these results to the more delicate case of
conductance based IF models 
in the present paper. (Note that  
having constant conductances leads to a dynamics which
is independent of the past firing times (raster plot). In fact, the dynamical
system is essentially a cellular automaton but with a highly non trivial dynamics).

\sssu{Alpha-profile conductances.}

Consider now a conductance based model of form (\ref{EqQCV}), leading to:

\beq \label{gammayvette}
\gamma_k(t,\tot) =K 
e^{
-\sum_{j} \bG_{kj}^\pm 
\left(\sum_{n=1}^{\mjt} \int_{t}^{t+1} \alpha^\pm(s-\tjn) \, ds 
\right)
},
\eeq

\nid where $K$ is a constant:

\beq\label{K}
K=e^{-\frac{\delta}{\tau_L}}<1.
\eeq

\nid while, using the form (\ref{alpha}) for $\alpha$ gives:

\beq\label{gammalphaprof}
\gamma_k(t,\tot)=K \, 
e^{
-\sum_{j=1}^N G_{kj} \sum_{n=1}^{\mjt} 
\left\{
\left[
\left(
1+\frac{1}{\tau} 
\right) \, e^{-\frac{1}{\tau}} - 1 
\right] 
- \frac{(t -\tjn)}{\tau} \, 
\left(
1-e^{-\frac{1}{\tau}}
\right)
\right\} \, e^{- \frac{t-\tjn}{\tau}}}.
\eeq

One has therefore to handle an exponential of an exponential. 
This example illustrates one of the main problem in  IF models.
IF models have been introduced to simplify neurons description
and to simplify numerical calculations (compared e.g. with Hodgkin-Huxley's model \cite{hodgkin-huxley:52}).
Indeed, their structure allows one to write an explicit expression for
the next firing times of each neurons, knowing the membrane potential
value. However, in case of $\alpha$  exponential profile, there is no
simple form for the integral and, even in the case of one neuron, one has to use approximations
with $\Gamma$ functions \cite{rudolph-destexhe:06} which reduce consequently
the interest of IF models and event based integration schemes.

\su{Theoretical analysis of the dynamics.}\label{y2dynamics}

\ssu{The general picture.} \label{SGen}

In this section we develop in words some important mathematical aspects
of the dynamical system (\ref{DNN}), mathematically proved in the sequel.\\

\textbf{Singularity set. }The first important property is that the dynamics (\ref{DNN}) (and the
dynamics of continuous time IF models as well)
 is not smooth, but has singularities, due to the sharp
threshold definition in neurons firing.
The singularity set is:
$$\cS=\left\{\V \in \cM | \exists i=1 \dots N, \mbox{\ such \ that} \
V_i=\theta \right\}.$$
This is the set of membrane potential vectors such that at least
one of the neurons has a membrane potential exactly equal to the threshold
\footnote{A sufficient condition for a neuron $i$ to fire at time $t$
is $V_i(t)=\theta$ hence $\V(t) \in \cS$. But this is not a necessary
condition. Indeed, as pointed in the footnote \ref{Notejump},
there may exist discontinuous jumps in the dynamics, even if time is continuous,
 either due to noise, or  $\alpha$ profiles with jumps (e.g. $\alpha(t) = \frac{1}{\tau}e^{-\frac{t}{\tau}}, \ t \geq 0$).
Thus neuron $i$ can fire with $V_i(t)>\theta$ and $\V(t) \notin \cS$.
 In the present case, this situation arises because time is discrete and
 one can have $V(t-\delta) < \theta$
and $V(t) >\theta$. 
This holds as well even if one uses  numerical
schemes using interpolations to locate more precisely the spike time \cite{hansel-et-al:98}.
}.
This set has a simple structure:
it is a finite union of $N-1$ dimensional hyperplanes. 
Although $\cS$ is a ``small'' set
both from the topological (non residual set)
 and metric  (zero Lebesgue measure) point of view,
it has an important effect on the dynamics.

\textbf{Local contraction.} The other important aspect is that the dynamics is locally \textit{contracting},
because $\gamma_k(t,\tot)<1$ (see eq. (\ref{gammacont})). 
This has the following  consequence. Let us consider the trajectory of a point $\V \in \cM$ and perturbations
with an amplitude $< \epsilon$ about $\V$ (this can be some fluctuation in the current,
or some additional noise, but it can also be some error due to a numerical implementation).
Equivalently, consider the 
evolution of the $\epsilon$-ball $\Bev$.
If $\Bev \cap \cS = \emptyset$ 
then we shall see in the next section that the image of $\Bev$ is a ball with a smaller
diameter. This means, that, under the condition $\Bev \cap \cS = \emptyset$, a perturbation
is \textit{damped}. Now, if the images of the ball under the dynamics never intersect
$\cS$,  any $\epsilon$-perturbation around $\V$ is exponentially damped
and the perturbed trajectories about $\V$ become asymptotically indistinguishable
from the trajectory of $\V$. Actually, there is a more dramatic effect. If all neurons have fired after a finite time $t$ then
all perturbed trajectories collapse onto the trajectory of $\V$ after $t+1$ iterations (see prop. \ref{PContract}
below). 

\begin{figure}[ht!]
\begin{center}
\includegraphics[height=5cm,width=5cm,clip=false]{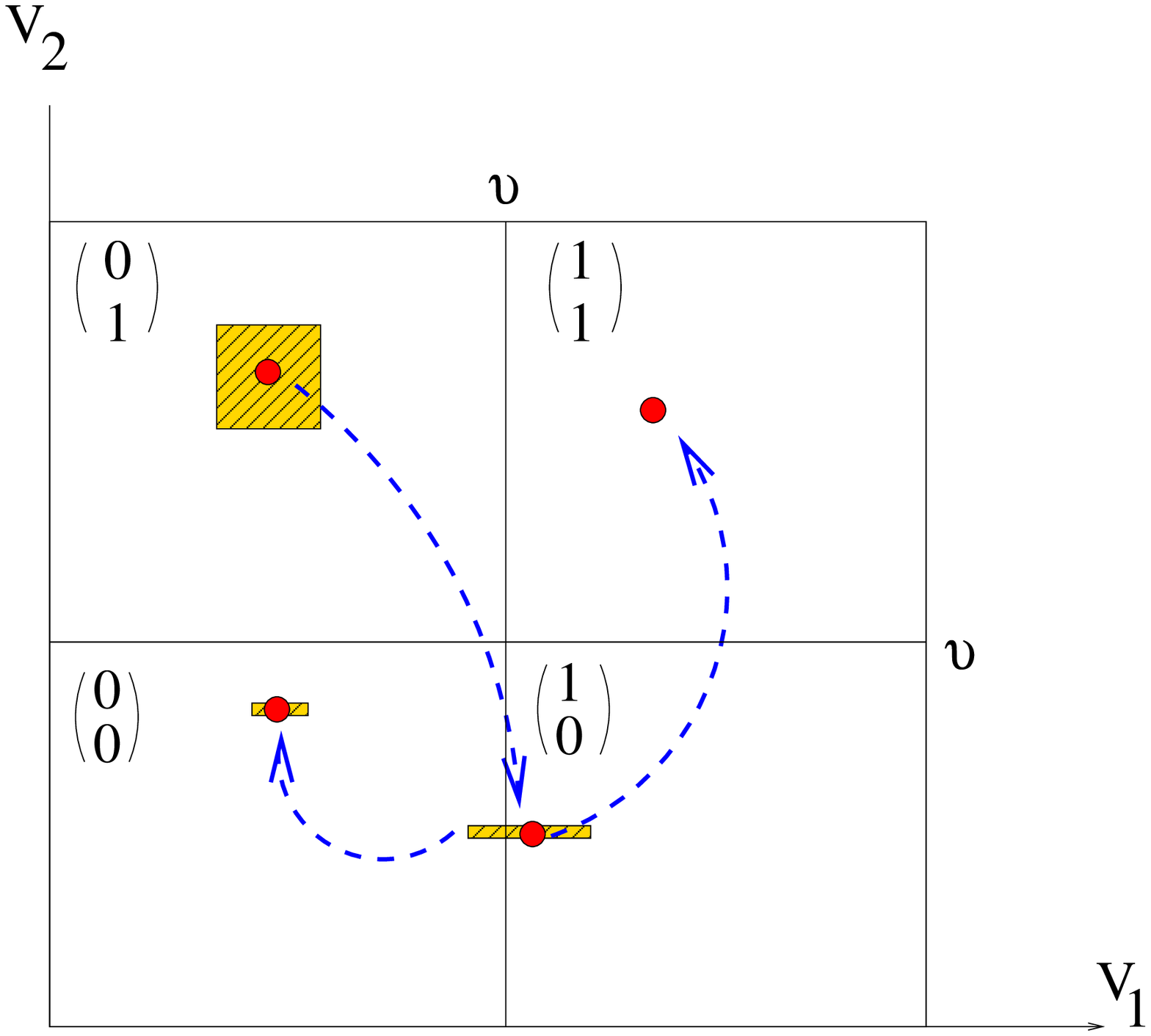}
\vspace{0.5cm}
\caption{
Schematic representation, for two neurons, of the natural partition $\cP$ and the mapping discussed in the text. 
In this case, a firing state is a vector with components
$\bom=\left(\stackrel{\omega_1}{\omega_2} \right)$ labeling the partition elements.
A set of initial conditions, say a small ($L^\infty$) 
ball in  $\cM_{\bom}$, is contracted by leak (neuron $1$ in the example) and reset (neuron $2$ in the example),
but its image can intersect the singularity set.
This generates several branches of trajectories. Note that we have given some width to the projection of
the image of the ball on direction $2$ in order to see it on the picture. But since neuron $2$ fires the
width is in fact $0$. 
}
\label{Sci}
\end{center}
\end{figure}

\textbf{Initial conditions sensitivity.} On the opposite, assume that there is a time, $t_0$, such that
the image of the ball $\Bev$ intersects  $\cS$.
By definition, this means that there exists a 
subset of neurons $\left\{i_1, \dots, i_k\right\}$ and
  $\V'  \in \Bev$, such that $Z(V_i(t_0))\neq Z(V'_i(t_0))$, 
$i \in \left\{i_1, \dots, i_k\right\}$. For example, some neuron does not fire
when not perturbed but the application of an $\epsilon$-perturbation induces
it to fire (possibly with a membrane
potential strictly above the threshold).  This requires obviously this neuron to be close enough to
the threshold. Clearly, the evolution of the unperturbed and perturbed trajectory
may then become drastically different (see Fig. \ref{Sci}). Indeed, even if only one neuron
is lead to fire when perturbed, it may induce other neurons to fire
at the next time step, etc \dots, inducing an avalanche phenomenon
leading to unpredictability  and initial condition sensitivity\footnote{This effect
 is well known in the context of synfire chains \cite{abeles:82,abeles:91,abeles-et-al:93,hertz:97}
or self-organized criticality \cite{blanchard-et-al:00}.
}.

It is tempting to call this behavior ``chaos'', but there is an important difference
with the usual notion of chaos in differentiable systems. In the present case,
due to the sharp condition defining the threshold,
initial condition only occurs at sporadic instants, whenever some neuron
is close enough to the threshold. 
Indeed, in certain periods of time the membrane potential typically is quite far below threshold, 
so that the neuron can fire only if it receives strong excitatory input over 
a short period of time. It shows then a behavior that is robust against fluctuations.
On the other hand, when membrane potential is close to the threshold
a small perturbation may induce drastic change in the evolution.

\textbf{Stability with respect to small perturbations.} Therefore,  depending on  parameters
such as the synaptic weights,
the external current, it may happen that, in the stationary regime, the 
typical trajectories stay away from the singularity set, say within a distance larger 
than $\epsilon >0$, which can be arbitrary small, (but positive). 
 Thus, a small perturbation (smaller than $\epsilon$) does not produce any
 change in the evolution.
At a computational level, this robustness leads to stable input-output transformations.
In this case, as we shall see, the dynamics of (\ref{DNN}) is asymptotically periodic
(but there may exist a large number of possible orbits, with a large period).
In this situation the system has a vanishing entropy\footnote{More precisely the topological
entropy (average bit rate production considered
over an \textit{infinite time horizon}) is zero but this implies that the Shannon entropy  is also zero.}.
This statement is made rigorous in theorem \ref{Tomega} below.

On the other hand, if the distance between the set where the asymptotic dynamics
lives\footnote{Say the ``attractor'', though one must be cautious with this notion, as
we shall see below.} and the singularity set
is arbitrary small then the dynamics exhibit initial conditions sensitivity,
and chaos. Thus, a natural question is: is chaos a generic situation ? How does
this depend on the parameters ? A related question is: how does the numerical
errors induced by a time discretization scheme evolve under dynamics \cite{hansel-et-al:98} ?

\textbf{``Edge of chaos''.} It has  been shown, in \cite{cessac:07} for the BMS model,
that there is a sharp transition\footnote{This transition is reminiscent of the one exhibited in  \cite{keener-et-al:81} for an isolated
neuron submitted to a periodic excitation, but the analysis
in \cite{cessac:07} and the present analysis hold  at the network level.
} from fixed point to complex dynamics, when crossing
a critical manifold usually called the ``edge of chaos'' in the literature. 
While this notion is usually not sharply defined in the Neural Network literature,
 we shall give a mathematical definition which is moreover tractable 
numerically.
Strictly speaking chaos only occurs on this manifold, but from a practical point
of view, the dynamics is indistinguishable\footnote{Namely,
though the dynamics is periodic, the periods are well
beyond the times numerically accessible.} from chaos, close to this manifold.
When the distance to the edge of chaos further increases the dynamics
is periodic with typical periods compatible with simulation times. This manifold
can be characterized in the case where the synaptic weights are independent, identically
distributed with a variance $\frac{\sigma^2}{N}$.

In BMS model (e.g., time discretized gIF model with constant conductances) it can be proved
that the chaotic situation is non generic \cite{cessac:07}. We now develop the same lines
of investigation  and discuss how these result extend to the model
(\ref{DNN}). Especially, the ``edge of chaos'' is numerically computed and compared to the BMS situation.

Let us now switch to the related mathematical results. Proofs are given in the appendix. 

\ssu{Piecewise affine map.} \label{SMap}

Let us first return to the notion of raster plot developed in section \ref{Rast}.
At time $t$, the firing
state $\bom(t) \in \Lambda$ can take at most $2^N$ values. Thus, the list of firing states
$\bom(0), \dots ,\bom(t) \in \Lambda^{t+1}$ can take 
at most $2^{N(t+1)}$ values. (In fact, 
as discussed below, only a  subset of these possibilities
is selected by the dynamics). This list is the raster plot up to time t
and  we have denoted by $\tot$. 
 Once the raster plot up to time $t$ has been fixed
the coefficients $\gamma_k$ and the integrated currents $J_k$ in (\ref{DNN}) are determined. 
Fixing the raster plot up to time $t$ amounts to construct branches for the discrete
flow of the dynamics, corresponding to sub-domains of $\cM$ constructed iteratively, via the natural partition (\ref{PartM}),
in the following way. \\

Fix $t > 0$ and $\tot$. Note:

$$\Mot=\left\{\V \in \cM | \V(s) \in \cM_{\bom(s), \ s= 0 \dots t} \right\}.$$
 
This is the set (possibly empty) of initial membrane potentials vectors $\V \equiv \V(0)$ whose firing
pattern at time $s$ is $\bom(s)$, $s=0 \dots t$. Consequently, $\forall \V \in \Mot$,
we have:

\beq\label{Vkt}
V_k(t+1)=\prod_{s=0}^{t} \gks\left[1-\omega_k(s)\right]V_k(0)
+\sum_{n=0}^{t} J_k(n,\ton) \prod_{s=n+1}^{t}  \gks\left[1-\omega_k(s)\right]  \ k=1 \dots N,
\eeq

\nid as easily found by recursion on (\ref{DNN}). We used the convention $ \prod_{s=t}^{t+1}  \gks\left[1-\omega_k(s)\right]=1$.

Then, define the map:

\beq\label{Ftot}
\Ftot = 
\left\{
\baR{lll}
\Mot &\to& \cM\\
\V & \to &  \Ftot\V=\V(t+1)
\eaR
\right.,
\eeq

\nid with $V_k(t+1)$ given by (\ref{Vkt})
and $\F_{\tom}^0=Id$. Note that $\Ftot$ is affine. Finally define:

\beq\label{Ft}
\left\{
\baR{ccc}
\F^{t+1} : \cM &\to& \cM\\
\V \in \Mot &\to& \FVTo	
\eaR
\right.
\eeq

\nid such that the restriction of $\F^{t+1}$ to the domain
$\Mot$ is precisely $\Ftot$.
This mapping is the flow of the model (\ref{DNN})
where:

$$\V(t+1) = \F^{t+1}\V, \quad \V \in \cM.$$

A central property of this map is that it is \textit{piecewise affine} and it has at most $2^{N(t+1)}$
branches $\Ftot$ parametrized by the legal sequences
$\tot$ which parametrize the possible histories
of the conductance/currents up to time $t$.\\

Let us give a bit of explanation of this construction. 
Take $\V \equiv \V(0) \in \cM_{\bom(0)}$. This amounts to fixing the firing pattern at time $0$
with the relation $\omega_k(0)=Z(V_k(0)), \ k=1 \dots N$. Therefore,
$V_k(1)=\gamma_k(0,\bom(0))\left[1-\omega_k(0)\right]V_k(0)+J_k(0,\bom(0))$,
where $\gamma_k,J_k$ do not depend on $\V(0)$ but only on the spike state
of neurons at time $0$. Therefore, the mapping $\F^1_{\tom} : \cM_{\bom(0)} \to \cM$
such that  $F^1_{k;\tom}\V=\gamma_k(0,\tom)\left[1-\omega_k(0)\right]V_k(0)+J_k(0,\tom), k=1 \dots N$
 is \textit{affine} (and continuous on the interior of $\cM_{\bom(0)}$). 
Since $\bom(0)$ is an hypercube, $\F^1_{\tom}\cM_{\bom(0)}$ is a convex connected domain.
This domain typically intersects several domains of the
natural partition $\cP$. This corresponds to the following situation.
Though the pattern of neuron firing at time $0$ is fixed as soon
as $\V(0) \in \cM_{\bom(0)}$, the list of neurons
firing at the next time depends on the \textit{value} of the membrane potentials $\V(0)$, and not
only on the spiking pattern at time $0$. But, by definition, the domain:

$$\cM_{\bom(1)\bom(0)} = \cM_{\left[\tom\right]_1}=
\left(
\F^{1}_{\tom}
\right)^{-1}\cM_{\bom(1)} \cap \cM_{\bom(0)}$$

\nid is such that $\forall \V(0) \in \cM_{\bom(1)\bom(0)}$,
the spiking pattern at time $0$ is $\bom(0)$ and it is $\bom(1)$
at time $1$. If the intersection is empty this simply means
that one cannot find a membrane potential vector such that
neurons fire according to the spiking pattern $\bom(0)$ at time
$t=0$ then fire according to the spiking pattern $\bom(1)$ at time
$t=1$. If the intersection is not empty we say that ``the transition
$\bom(0) \to \bom(1)$ is \textit{legal}\footnote{Conditions ensuring
that a transition is legal depend on the parameters of the dynamical
systems, such as the synaptic weights.}''. 

Proceeding recursively as above one constructs a hierarchy
of domains $\Mot$ and maps $\Ftot$. Incidentally, an equivalent
definition of $\Mot$ is:

\beq\label{Mot}
\Mot=\bigcap_{s=0}^{t} \left(\F^s_{\tom} \right)^{-1}\cM_{\bom(s)}.
\eeq

As stated before,  $\Mot$ is the set of membrane potential vectors $\V$
such that the firing patterns up to time $t$ are $\bom(0), \dots, \bom(t)$.
If this set is non empty we say that the sequence $\bom(0), \dots, \bom(t)$
is \textit{legal}. Though there are at most $2^{N(t+1)}$ possible raster plots at time $t$
the number of legal raster plots is typically smaller. This number can increase either
exponentially with $t$ or slower.  
We shall denote by $\SLp$ the set of all legal (infinite) raster plots (legal infinite sequences of firing
states). Note that $\SLp$ is a topological space for the product topology
generated by cylinder sets \cite{Katok-Hasselblatt:98}. The set $\tot$ of raster plots
 having the same first $t+1$ firing patterns
is a cylinder set.

\ssu{Phase space contraction.} \label{SContract} 

Now, we have the following:

\bp \label{PContract}

For all $\tom \in \SLp, \forall t \geq 0$,  the mapping $\V \in \Mot \to 
\FVTo$ is affine, with a Jacobian matrix and an affine constant depending
on $t,\tot$. Moreover, the Jacobian matrix  is diagonal
with eigenvalues 

$$\sigma_k(t,\tom)=\prod_{s=0}^t \gamma_k(s,\tos)(1-\omega_k(s))<1, k=1 \dots N.$$

\nid Consequently, $\FVTo$ is a contraction.
\ep

\bpr The proof results directly from the definition
(\ref{Ftot}) and (\ref{Vkt}) with $\gamma_k(s,\tos)<1, \forall s \geq 0$ (see (\ref{gammacont}) ). 
\epr

Since the domains $\cMo$ of the natural partition are convex and connected,
and since $\F$ is affine on each domain (therefore continuous on its interior),
there is a straightforward corollary:

\bcor\label{CConv}
The domains $\Mot$ are convex and connected.
\ecor

There is a more important, but still straightforward consequence:

\bcor\label{ContFt}
$\F^{t+1}$ is a non uniform contraction on $\cM$ where the contraction
rate in direction $k$ is  
$\frac{1}{t+1} \sum_{s=0}^{t} \log\left[\sigma_k(s,\tom)\right]$,
$\forall \V \in \Mot$.
\ecor

Then, we have the following :

\bp\label{CCont}
Fix $\tom \in \SLp$. 

\ben
\item If  $\exists t <\infty$,  such that, $\forall k=1 \dots N$,
 $\exists s\equiv s(k) \leq t$ where 
$\omega_k(s)=1$ then $\Ftot[\Mot]$ is a \underline{point}. That is,
all orbits born from the domain $\Mot$ converge
to the same orbit in a finite time. 

\item If  $\exists k \in \left\{1, \dots N\right\}$ such that
$\forall t>0$, $\omega_k(t)=0$ then $\Ftot$ is contracting in direction
$k$, with a Lyapunov exponent $\lambda_k(\tom)$, such that:

$$\liminf_{t \to \infty} \frac{1}{t+1} \sum_{s=0}^{t} \log \gamma_k(s,\tos)
\leq \lambda_k(\tom) \leq  \limsup_{t \to \infty} \frac{1}{t+1} \sum_{s=0}^{t} \log \gamma_k(s,\tos) <0$$
\een
\ep

\bpr
Statement 1 holds since, under these hypothesis, all eigenvalues of $\Ftot$ are zero. 
For 2, since $D\Ftot$ is diagonal, the Lyapunov exponent in direction $k$ is defined by 
$\lambda_k(\tom)=\lim_{t \to \infty}
\frac{1}{t+1} \sum_{s=0}^{t} \log(\sigma_k(t,\tom))$ whenever the limit exists (it exists almost surely
for any $\F$ invariant measure from Birkoff theorem).
\epr
 
\textbf{Remark.} An alternative definition of Lyapunov exponent has been introduced by 
Coombes in \cite{coombes:99} and \cite{coombes:99b},
for Integrate and Fire neurons. His definition, based on ideas developed for impact
oscillators \cite{muller:95}, takes care of the discontinuity in the trajectories arising when crossing
$\cS$. Unfortunately, his explicit computation  at the network level (with continuous time dynamics), 
 makes several implicit assumptions (see eq. (6) in \cite{coombes:99b}): (i) there is a finite number of spikes within
a bounded time interval; (ii) the number of spikes that have been fired up to time $t$, 
$\forall t>0$, is the same for
the mother trajectory and for a daughter trajectory, generated
by a small perturbation of the mother trajectory at $t=0$;
(iii) call $T_i^k$, in Coombes' notations, the $k$-th spike time for neuron $i$
in the mother trajectory, and $\tilde{T}_i^k$ the $k$-th spike time for neuron $i$  in the daughter trajectory.
Then  $\tilde{T}_i^k=T_i^k + \delta_i^k$, where $\delta_i^k$ is assumed to become arbitrary small, $\forall k \geq 0$,
when the initial perturbation amplitude tends to zero.
While assumption (i) can be easily fulfilled (e.g. by adding a refractory period)
assumptions (ii) and (iii) are more delicate. 

Transposing this computation to the present analysis,
this requires that both trajectories are never separated by the singularity set. 
A sufficient condition is that the mother trajectory stays sufficiently far from the singularity set.
In this case the Lyapunov exponent defined by Coombes coincides with our definition
and it is negative. On the other hand, in the ``chaotic'' situation (see section \ref{ghost}), assumptions
(ii) and (iii) can typically fail. For example, 
it is possible that neuron $i$ stops firing after a certain time, in the daughter trajectory,
while it was firing in the mother trajectory, and this can happen even if the perturbation is 
arbitrary small. This essentially means that
the explicit formula for the Lyapunov exponent proposed in  \cite{coombes:99b} cannot be applied
as well in the ``chaotic'' regime.

\ssu{Asymptotic dynamics.}\label{SAsdyn}

\sssu{Attracting set $\cA$ and $\omega$-limit set.}\label{SAtt}

The main notion that we shall be interested in from now on concerns
the invariant set where the asymptotic
dynamics lives.

\bdf  (From \cite{guckenheimer-holmes:83,Katok-Hasselblatt:98})
A point $y \in \cM$ is called an $\omega$-limit point
for a point $x \in \cM$ if there exists a sequence of times
$\left\{t_k\right\}_{k=0}^{+\infty}$, such that
$x(t_k) \to y$ as  $t_k \to +\infty$. The $\omega$-limit set of $x$, $\omega(x)$,
is the set of all $\omega$-limit points of $x$. The $\omega$-limit
set of $\cM$, denoted by $\oM$, is the set $\oM = \bigcup_{x \in \cM}\omega(x)$.
\edf

Equivalently, since $\cM$ is closed and invariant, we have
$\oM=\bigcap_{t=0}^\infty \F^t(\cM)$.

Basically, $\Omega$ is the union of attractors.
But for technical reasons, related to the case 
considered in section \ref{ghost},
it is more convenient to use the notion
of $\omega$-limit set.

\sssu{A theorem about the structure of $\oM$.} \label{SStructA}

\bth\label{Tomega} 
Assume that $\exists \epsilon >0$ and $ \exists T < \infty$ such that,
$\forall \V \in \cM$, $\forall i \in \left\{1 \dots N \right\}$,
\ben
\item Either $\exists t \leq T$ such that $V_i(t) \geq  \theta$;
\item Or $\exists t_0 \equiv t_0(\V,\epsilon)$
 such that $\forall t \geq t_0$, $V_i(t) < \theta -\epsilon$
\een
Then,  $\oM$ is composed by finitely many
periodic orbits with a period $\leq T$. 
\enth 

The proof is given in the appendix \ref{Pomega}.\\

Note that conditions (1) and (2) are not disjoint. The meaning
of these conditions is the following. (1) corresponds to
imposing that a neuron has fired after a finite time $T$
(uniformly bounded, i.e. independent of $\V$ and $i$).
(2) amounts to requiring that  after a certain
time $t_0$, the membrane potential stays  below the threshold value 
and it cannot accumulate on $\theta$. We essentially want to avoid
a situation when a neuron can fire for the first time after
an unbounded time (see section \ref{ghost} for a discussion
of this case).
Thus assumptions (1),(2) look quite reasonable.
 Under these assumptions the asymptotic dynamics
is periodic and one can predict the evolution after observing the system
on a finite time horizon $T$, whatever the initial condition.
Note however that $T$ can be quite a bit large.

There is a remarkable corollary result, somehow hidden in the proof given in the appendix.
The neurons that do not fire after a finite time are still driven by the dynamics of
firing neurons. It results  that,
in the asymptotic regime, non firing neurons have a membrane potential which
oscillates below the threshold. This exactly corresponds to what people call
``sub-threshold oscillations''. In particular, there are times where those
membrane potentials are very close to the threshold, and a small perturbation
can completely changes further evolution. This issue is developed in the next section.
Possible biological interpretations are presented in the discussion section.

\sssu{Ghost orbits.}\label{ghost}

The advantage of the previous theorem is that we define conditions 
where one can relatively easily controls dynamics. However, what happens
if if we consider the complementary situation corresponding to the following
definition ?

\bdf\label{DfGhost}
An orbit $\tV$ is a \textit{ghost orbit} if  $\exists i$ such that:

$$(i) \; \forall t> 0, V_i(t)<\theta$$ 

\nid and :

$$(ii) \; \limsup_{t \to +\infty} V_i(t)=\theta$$
\edf

 Namely there exists
at least one initial condition $\V$ and one neuron $i$ such that  
one cannot uniformly bound the first time of firing, and $V_i(t)$ approaches
arbitrary close the threshold. In other words sub-threshold oscillations drive 
the neuron ``dangerously close'' to the threshold, though we are not able to
predict when the neuron will fire.
This may look a ``strange'' case from a practical point of view,
but it has deep implications. 
This indeed means that we can observe the dynamics on arbitrary long times
without being able to predict what will happen later on, because when
this neuron eventually fire, it may drastically change
the evolution. This case is exactly related to the chaotic
or unpredictable regime of IF models.

One may wonder whether
the existence of ghost orbits is ``generic''.
To reply to this question one has first to give a definition
of genericity. In the present context, it is natural
to consider the dynamical system describing the time evolution
of our neural network as a point in a space $\cH$ of parameters.
These parameters can be e.g., the synaptic weights, or parameters
fixing the time scales, the reversal  potentials, the
 external currents, etc... 
Varying these parameters (i.e., moving the point representing our
dynamical system in $\cH$) can have two possible effects.
Either there is no qualitative change in the dynamics and observable
quantities such as e.g., firing rates, average inter-spikes interval, etc,  are varying continuously. Or,
a sharp change (bifurcation) occurs. This corresponds to the crossing
of a critical or bifurcation manifold in $\cH$.
Now, a behavior is generic if it is ``typical''. On mathematical
grounds this can have two meanings. Either this behavior is
obtained, with a positive probability, when drawing the parameters (the corresponding point in $\cH$)
at random with some natural probability (e.g., Gaussian). In this case one speaks of ``metric
genericity''. Or, this behavior holds in a dense subset of $\cH$.
One speaks then  of ``topological genericity''. The two notions usually
do not coincide. 
 
In the BMS model, ghost orbits are non generic in both senses \cite{cessac:07}.
 The proof does not extend to more general models such as (\ref{DNN}) because it heavily uses the fact that the synaptic current
takes only finitely many values in the BMS model. As soon as we introduce a dependence in $\tom$ this is not
the case anymore. We don't know yet how to extend this proof. 

\ssu{Edge of chaos.}\label{edge}

On practical grounds ghost orbits involve a notion
of limit $t \to +\infty$ which has no empirical meaning. Therefore the right question
is: are there situations where
a neuron can fire for the first time after
a time which is well beyond the observation
time ? One way to analyze this effect is to consider
how close the neurons are to the threshold
in their evolution. On mathematical
grounds this is given by the distance
from the singularity set to the $\omega$-limit set:

\beq\label{dAS}
\dAS= \inf_{\V \in \oM}\inf_{t \geq  0} \min_{i =1 \dots N} |V_i(t) - \theta|.
\eeq

The advantage of this definition, is that it can easily be adapted to 
the plausible case where observation time is bounded (see section \ref{y2num}).\\

Now, the following theorem holds.

\bth.\label{ThdAS}

\ben

\item If $\dAS>0$ then $\Omega$ is composed by finitely many periodic orbits with a finite period.

\item There is a one-to-one correspondence between a trajectory on $\Omega$ and its raster plot.

\een

\enth

The proof is exactly the same as in \cite{cessac:07} so we don't reproduce it here.
It uses the fact that if $\dAS>0$ then there is a finite time $T$, depending on $\dAS$
and diverging as $\dAS \to 0$, such that $\F^{T}$
has a Markov partition (constituted by local stable manifolds since dynamics
is contracting) where the elements of the partition are the domains $\MoT$. 
Note however that $ \dAS>0$ is a sufficient but not a necessary condition to have a periodic
dynamics. In particular, according to theorem \ref{Tomega} one can have $\dAS=0$
and still have a periodic dynamics, if at some finite time $t$, for some neuron $i$, $V_i(t)=\theta$.
This \textit{strict} equality is however not structurally stable, since a slight change e.g. in the
parameters  will remove it. The main role of the condition $\dAS>0$ is therefore 
to avoid situations
where the membrane potential of some neuron accumulates on $\theta$ \textit{from below} (ghost orbits).
See the discussion section for a possible biological interpretation on this. 

But $\dAS$ plays another important role concerning the effect of perturbations
on the dynamics. Indeed, if $\dAS>0$ then the contraction property (corollary \ref{ContFt}) implies
that any perturbation smaller than $\dAS$ will be damped by dynamics.
In particular, making a small error on the spike time, or any other
type of error leading to an indeterminacy of $V$ smaller than $\dAS$
will be harmless.

Let us now discuss  the second item of theorem \ref{ThdAS}. It expresses that the raster plot is a \textit{symbolic
coding} for the membrane potential trajectory. In other words there is no loss
of information on the dynamics when switching from the membrane potential
description to the raster plot description. This is not true anymore if $\dAS=0$.

The first item tells us that dynamics is periodic, but period can be  arbitrary long. Indeed,
 following  \cite{cessac:07} an estimate for an upper bound on the orbits
period is given by:

\beq \label{nM}
n_M \simeq 2^{N\frac{\log(\dAS)}{\log(<\gamma>)}}
\eeq

\nid where $<\gamma>$ denotes the value of $\gamma$ averaged over time and initial
conditions\footnote{Note that the system is not uniquely ergodic (see \cite{Katok-Hasselblatt:98} for a definition of unique ergodicity).} (see appendix for details).
Though this is only an upper bound this suggests that periods diverge when $\dAS \to 0$.
In BMS model, this is consistent with the fact that when $\dAS$ is close to 0 dynamics ``looks chaotic''.
Therefore, $\dAS$ is  what a physicist could call an ``order parameter'',
quantifying somehow the dynamics complexity.
The distance $ \dAS$ can be numerically estimated as done in (\ref{num-def-1}) and (\ref{num-def-2}), section \ref{y2num}.\\

Before, we need the following list of (operational) definitions.

\bdf (Edge of chaos)

The edge of chaos is the set of points $\cE$ in the parameter space $\cH$ where
$\dAS=0$.
\edf

The topological structure of $\cE$ can be quite complicated as
we checked in the simplest examples (e.g., the BMS model with Laplacian
coupling) and suggested by the papers \cite{bressloff-coombes:00,bressloff-coombes:00b}
(see discussion). There are good reasons to believe
that this set coincides with the set of points where
the entropy is positive (see \cite{kruglikov-rypdal:05,kruglikov-rypdal:05b} and discussion below).
The set of points where the entropy  is positive can  have a fractal
structure even in the simplest examples of one dimensional
maps \cite{mackay-tresser:86,gambaudo-tresser:88}. Therefore, there is no hope
to characterize $\cE$ rigorously in a next future. Instead,
we use below a numerical characterization.

The edge of chaos is a non generic set in the BMS model, and 
the same could hold as well in model (\ref{DNN}).
Nevertheless, it has a strong influence on the dynamics,
since crossing it leads to drastic dynamical changes.
Moreover, close to $\cE$ dynamics
can be operationally indistinguishable from chaos.
More precisely, let us now propose another definition.

\bdf (Effective entropy)

Fix $T_o$ called ``the observational time''. This is the largest accessible duration of
an experiment. Call $n(t)$ the number of (legal) truncated raster plots up to time
$t$. Then, the effective entropy is;

\beq\label{entro}
\heff=\frac{1}{T_o}\log{n(T_o)}
\eeq

\edf

Note that in the cases where raster plots provide a symbolic coding for the
dynamics then $\lim_{T_o \to \infty} \heff=h^{(top)}$,
the topological entropy.\\

On practical grounds, this definition corresponds to the following notion.
The larger the effective entropy, the more the system is able to produce
distinct \textit{neural codes}. This provides one way to measure the ``complexity''
of the dynamics. On more ``neuronal'' grounds this quantity measures 
the variability in the dynamical response of the neuronal network to a given 
stimulus (external current) or its ability to produce distinct ``functions'' (a function
being the response to a stimulus in terms of a spikes train). Thinking of learning mechanisms (e.g., Hebbian) 
and synaptic plasticity (LTD,LTP,STDP) one may expect to having the largest
learning capacities when this entropy is large. This aspect will be developed
in a separated paper. (For the effect of Hebbian learning and entropy reduction
in firing rate neural networks see \cite{Siri:07a}).

Finally, a positive effective entropy means  that the system
essentially behaves like a chaotic system during the time
of the experiment. Indeed, the entropy is closely related to
the distance $\dAS$, since, from (\ref{nM}), a rough estimate/bound of $\heff$ is easily obtained from
(\ref{nM}), (\ref{entro}):

\beq
\heff < N\frac{\log(\dAS)}{T_o\log(<\gamma>)}\log(2) \label{boundh}
\eeq

The notion of  effective entropy provides
 some notion of ``width'' to the edge of chaos $\cE$.
For a fixed $T_o$ the system behaves chaotically
in a thick region $\cE_{T_o} \supset \cE$ in $\cH$ such that $\heff>0$. 
And from (\ref{boundh}) one expects that this entropy gets larger when $\dAS$ gets smaller.

\ssu{Effects of time discretization.}\label{RqTdisc}

Under the light of the previous results, let us reconsider the approximation where spikes
are taken into account at multiple of the characteristic time scale $\delta$, for the conductances update.
Doing this, we make some error in the computation of the membrane potential,
compared to the value obtained when using the ``exact'' spike time value.
Now, the question is whether this error will be amplified by the dynamics,
or damped. As we saw, dynamics (\ref{DNN}) is contracting but the effect
of a small error can have dramatic consequences when approaching the singularity
set. The distance $\dAS$ provides a criterion to define a ``safe'' region
where a small error of amplitude $\epsilon>0$ in the membrane potential  value is harmless,
basically, if $\epsilon < \dAS$. On the other hand, if we are in a region of the parameters space where $\dAS=0$
then a slight perturbation has an incidence on the further evolution. 
Since $\delta$ can be arbitrary small
in our theorems we have a good  control on the dynamics of the continuous
time IF models
 \textit{except at the edge of chaos} where $\dAS=0$.
This enhances the question of \textit{mathematically}  characterizing
this region in the parameter space $\cH$. Note indeed that numerical
investigations are of little help here since we are looking
for a parameter region where the distance $\dAS$ defined as an \textit{asymptotic}
limit, has to be \textit{exactly} zero.
 The problem is that even sophisticated schemes (e.g., event based)
are also submitted to round off errors. Therefore, as a conclusion, it might well be that \textit{all} 
numerical schemes designed to approximate continuous time Integrate and Fire models
 display trajectory sensitivity to spike time errors, when approaching $\dAS=0$.

\su{A numerical characterization of the ``edge of chaos''.}\label{y2num}

 \ssu{A ``coarse-grained'' characterization.}

As mentioned in the previous section an exact 
analytic computation of the edge of chaos
in the general case seems to be out of reach.
However, a ``coarse grained'' characterization
can be performed at the numerical level and possibly
some analytic approximation could be obtained. For this, we choose  
the synaptic weights (resp. the synaptic conductances) (and/or the external currents) randomly, with some
probability $\PrW$ ($\PrI$), where $\cW$ is the matrix of synaptic weights
($W_{ij}=E^{\pm}G_{ij}^{\pm}$)
and $\Ie$ the vector of external currents (recall that external currents are
time constant in this paper).
A natural  starting point is the use of Gaussian independent, identically
distributed variables, where one
varies the parameters, mean and variance, defining the probability definition (we call them
statistical parameters, see \cite{cessac-samuelides:07,samuelides-cessac:07} for further
developments on this approach). Doing these,
one performs a kind of fuzzy sampling of the parameters space, and one somehow expects
the behavior observed for a given value of the statistical parameters to be characteristic
of the region of $\cW,\Ie$ that the probabilities $\PrW,\PrI$ weight (more precisely, one expects to observe
a ``prevalent'' behavior in the sense of \cite{sauer:92}). 

The idea is then to estimate numerically $\dAS$ in order to characterize how it varies
when changing the statistical parameters. As an example, in the present paper,
 we select conductances (resp. synaptic weights)
randomly with a  Gaussian probability with
a fixed mean and a variance $\frac{\sigma^2}{N}$, and we study the behavior
of $\dAS$
when $\sigma$ is varying.
Note that the neural network is almost surely fully connected. 
We compute numerically an approximation of the distance $\dAS$, where we fix
a transient time $T_r$ and an observation time $T_o$ and average over several initial conditions $\V^{(n)}, n=1 \dots N_{CI}$
for a given sample of synaptic weights. Then we perform an average over
several synaptic weights samples $\cW_m, m= 1 \dots N_{\cW}$. In a more compact form, we compute:  

\beq \label{num-def-1}
\dASe=\frac{1}{N_{\cW}} \sum_{m=1}^{N_{\cW}} d_{\cW_m}^{(exp)},
\eeq

\nid where:

\beq \label{num-def-2}
d_{\cW_m}^{(exp)}  = \min_{\V^{(n)}, n= 1 \dots N_{CI} }
\min_{t = 1 \dots T_o}
\min_{i =1 \dots N} |V^{(n)}_i(T_r+t) - \theta|.
\eeq

In this way, we obtain a rough and coarse grained
location of the edge of chaos where the distance $\dAS$
vanishes. \\

We have performed the following experiments with two control parameters.

\bit
\item\textbf{Variance of random synaptic weights.} We randomly select the synaptic strength 
which modulates the synaptic conductance using a Gaussian distribution so that 80\% of 
the synapses are excitatory and 20\% inhibitory. The average standard-variation $\sigma$ is varied.
The value $\sigma=0.5$ corresponds, in our units, 
to what is chosen in the literature when considering the biological dispersion in the cortex (e.g., \cite{rudolph-destexhe:06}).
Note however that, at the present stage of investigation, Dale's principle  is 
not taken into account.

\item\textbf{Membrane leak time-constant.} As an additional control parameter 
we vary the membrane leak around the usual $\tau_L = 1 \cdots 20 ms$ values. This choice is two-fold. The value $\tau_L = 20 ms$ corresponds to 
in-vitro measurement, while $\tau_L \rightarrow 1 ms$ allows to represent in-vivo conditions in the cortex. 
On the other hand, it acts directly on the 
 average contraction rate $<\gamma>$ which is a natural control parameter.

\eit

 Each simulation randomly selects the initial potential values in a $-70 \dots 30mV$ range. 
For each condition the simulation is run for $N_{CI}=100$ initial conditions and $N_{\cW}=10$ random
selection of the synaptic weights. With a sampling period of $0.1ms$, the network is run during $T_r=1000$ steps in order to ``skip'' the transients
\footnote{Note the transients depend on the parameters and on the distance to
the singularity set too. In particular, one can have transients that are well beyond
the current capacity of existing computers. Therefore, our procedure gives
a rough localization of the edge of chaos. Analytic computation
would give a more precise localization. 
} and 
then observed during $T_o=1000$ steps. In order to explore the role of history dependent conductances on the dynamics we considered different
models from the biologically plausible IF model to BMS model. 
More precisely we explore four modeling variants: 

\ben

\item\textit{Model I, defined in (\ref{DNN})}.

\item\textit{Model II (\ref{DNN}) with a fixed $\gamma$.} The evolution equation of membrane potentials is thus given by:

$$V_k(t+1) = \left<\gamma\right> V_k(t) \, [1 - Z(V_k(t))] + J_k(t, \tot)$$

\nid
where the average $\left<\gamma\right>$ is the  value observed during the 
numerical simulation of model I.
 Note that $\left<\gamma\right>$ depends on the parameters $\sigma,\tau_L$.
The goal of this simulation is to check the role of the fluctuations of $\gamma(t,\tot)$, controlling
the instantaneous contraction rate, compared to a mean-field model where $\gamma(t,\tot)$ is replaced by its average.
This corresponds to what is called ``current based'' synapses instead of ``conductance based'' synapses in the 
literature, (see e.g. \cite{brette-rudolph-etal:07}).

\item\textit{Model III (\ref{DNN}) approximation with a fixed $\gamma$ and simplified synapses.} The evolution equation of membrane potentials is given by:

$$V_k(t+1) =\left<\gamma\right>V_k(t) \, [1 - Z(V_k(t))] 
+ E^+ \sum_{j} \, G_{ij}^+ \, Z(V_j(t - \delta^+))
+ E^- \sum_{j} \, G_{ij}^- \, Z(V_j(t - \delta^-))
$$

In addition to the previous  simplification, we consider 
so-called ``current jump'' synapses where the synaptic input 
simply corresponds to an impulse, added to the membrane potential equation.
Here the magnitude of the impulse and its delay $\delta^- = 2ms$ and $\delta^+ = 10ms$
in order to keep both characteristics as closed as possible to the previous condition.

\item\textit{Model IV (\ref{DNN}) with a fixed $\gamma$ and instantaneous simplified synapses.}
  The evolution equation of membrane potentials is given by:

$$V_k(t+1) = <\gamma> \, V_k(t) \, [1 - Z(V_k(t))] +\sum_{j} W_{ij} Z(V_j(t)))$$

\nid where in addition to the previous simplification,
the delay has been suppressed and where $W_{ij}=E^\pm G_{ij}^\pm$. This last condition strictly correspond to the original BMS model \cite{soula-beslon-etal:06}.

\een

\begin{figure}[ht!]
\begin{center}
\includegraphics[height=7cm,width=7cm,clip=false]{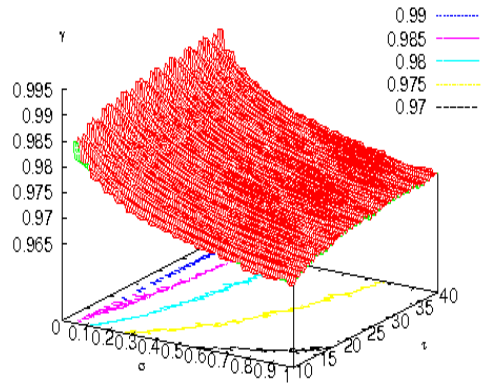}

\vspace{0.5cm}
\caption{Average value of $\gamma$ for model I  $\sigma \in [0.01,1]$, $\tau_L \in [10,40] ms$. The profile is very similar for other models.}
\label{Fgamm}\end{center}
\end{figure}

The results are given below.
We have first represented the average value $\left<\gamma\right>$ for the model I
in the range  $\sigma \in [0.01,1]$, $\tau_L \in [10,40] ms$ (see Fig. \ref{Fgamm}).
The quantity related to the contraction rate, is remarkably constant (with small variations within the range $[0.965,0.995]$).

%
%%%%%%%%%%%%%%%%%%%% Cas 1cdf
%%%%%%%%%%%%%%%%%%%%%%%%%

\begin{figure}[ht!]
\begin{center}
\includegraphics[height=7cm,width=7cm,clip=false]{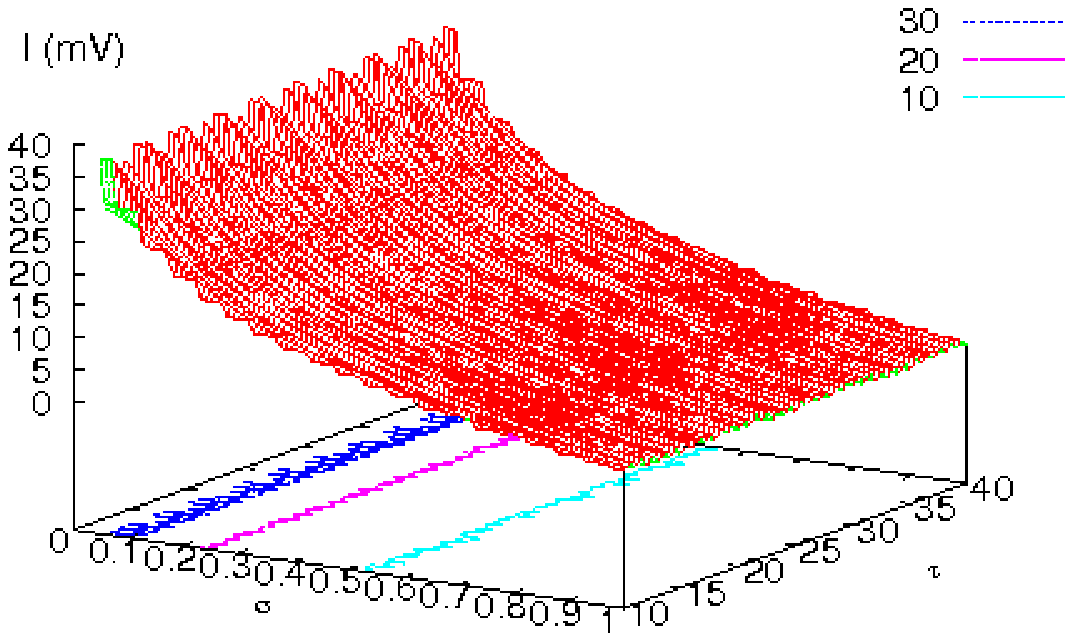}
\includegraphics[height=7cm,width=7cm,clip=false]{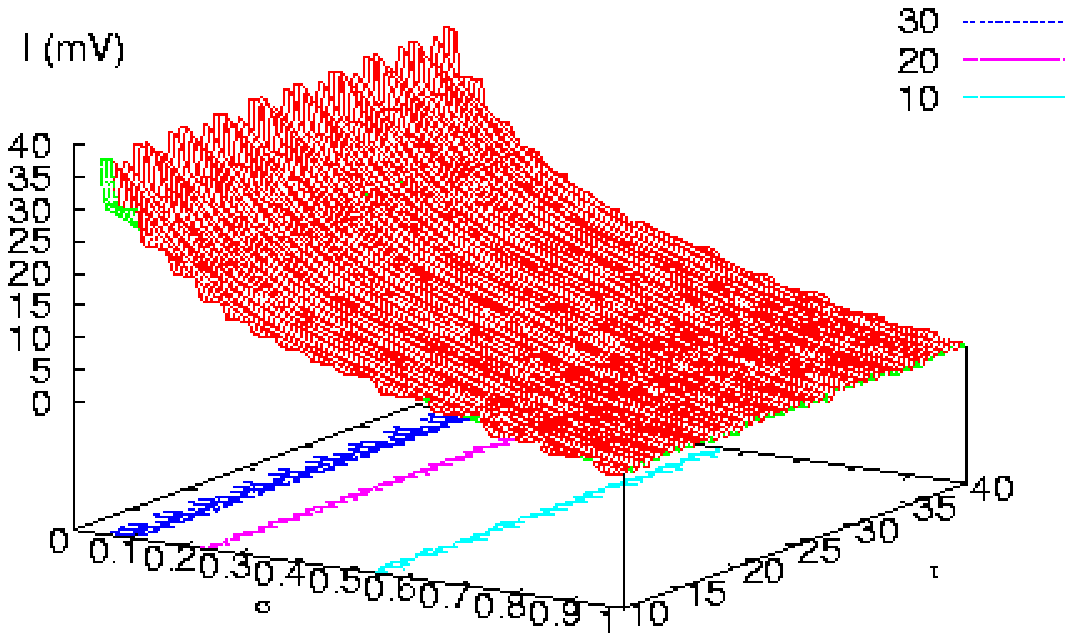}

\vspace{0.5cm}

\includegraphics[height=7cm,width=7cm,clip=false]{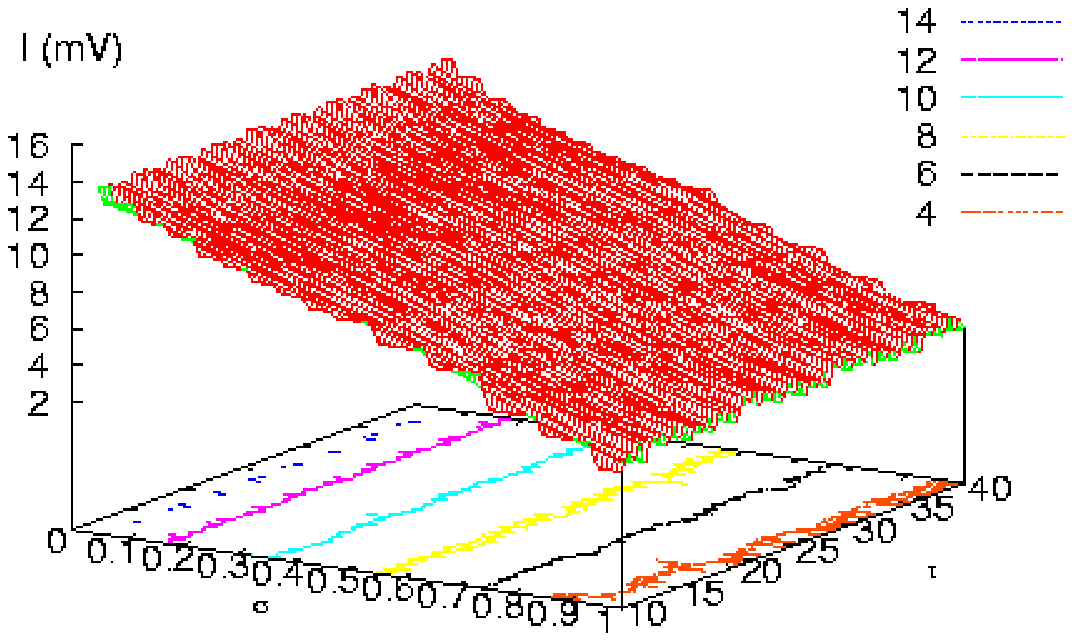}
\includegraphics[height=7cm,width=7cm,clip=false]{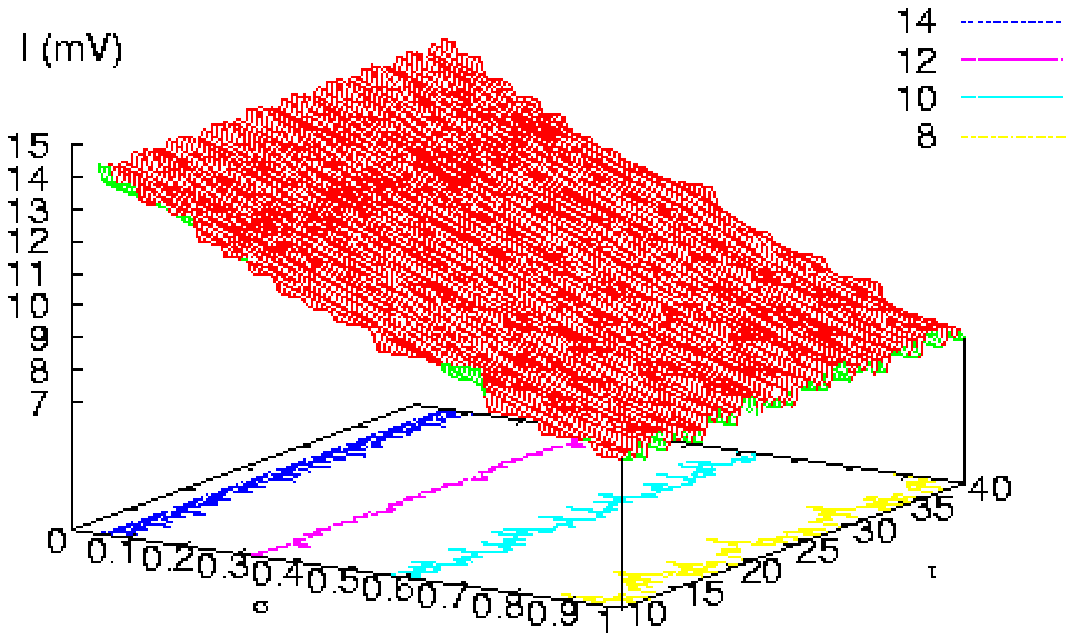}

\vspace{0.5cm}
\caption{Average current $I$ for the models I (top left)-II (top right)- III (bottom left)-
IV (bottom right)  with $\sigma \in [0.01,1]$, $\tau_L \in [10,40] ms$. }
\label{Fcurr}\end{center}
\end{figure}

\begin{figure}[ht!]
\begin{center}
\includegraphics[height=7cm,width=7cm,clip=false]{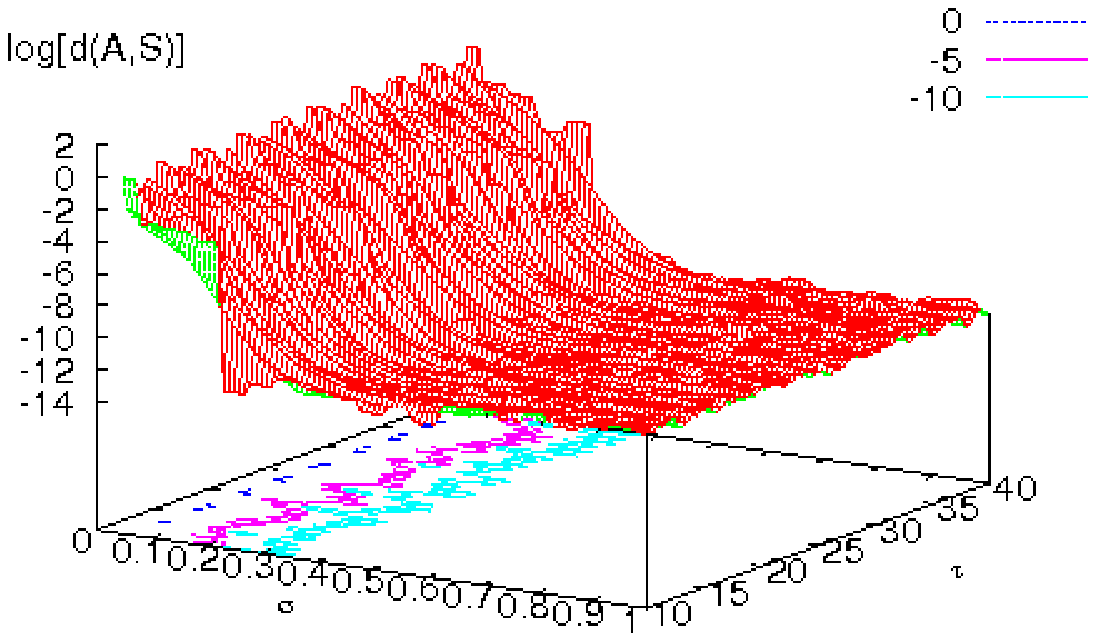}
\includegraphics[height=7cm,width=7cm,clip=false]{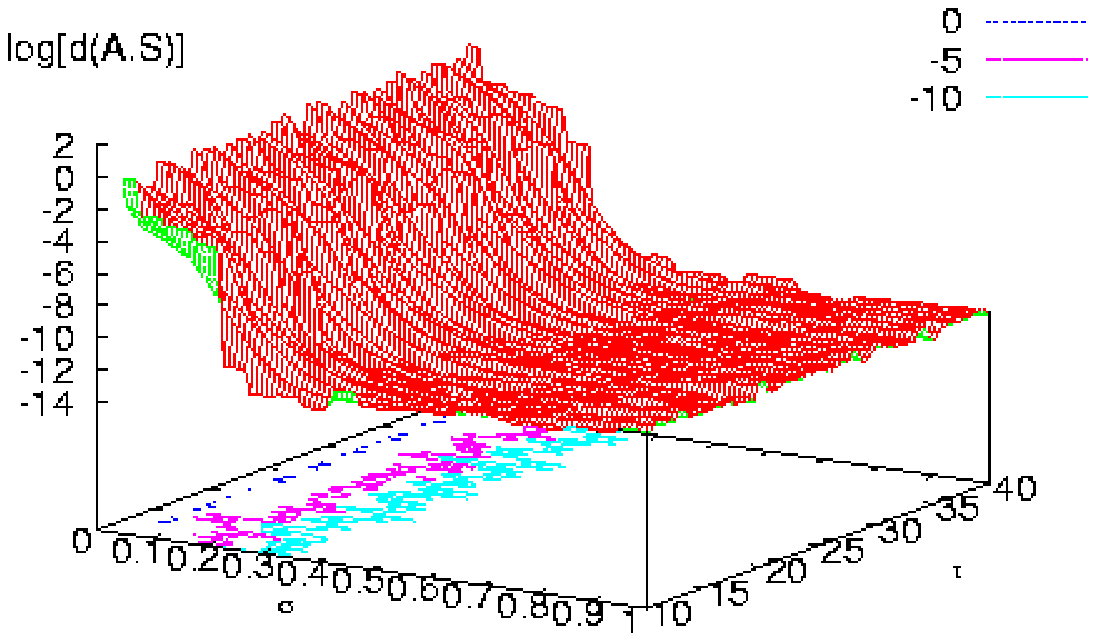}

\vspace{0.5cm}

\includegraphics[height=7cm,width=7cm,clip=false]{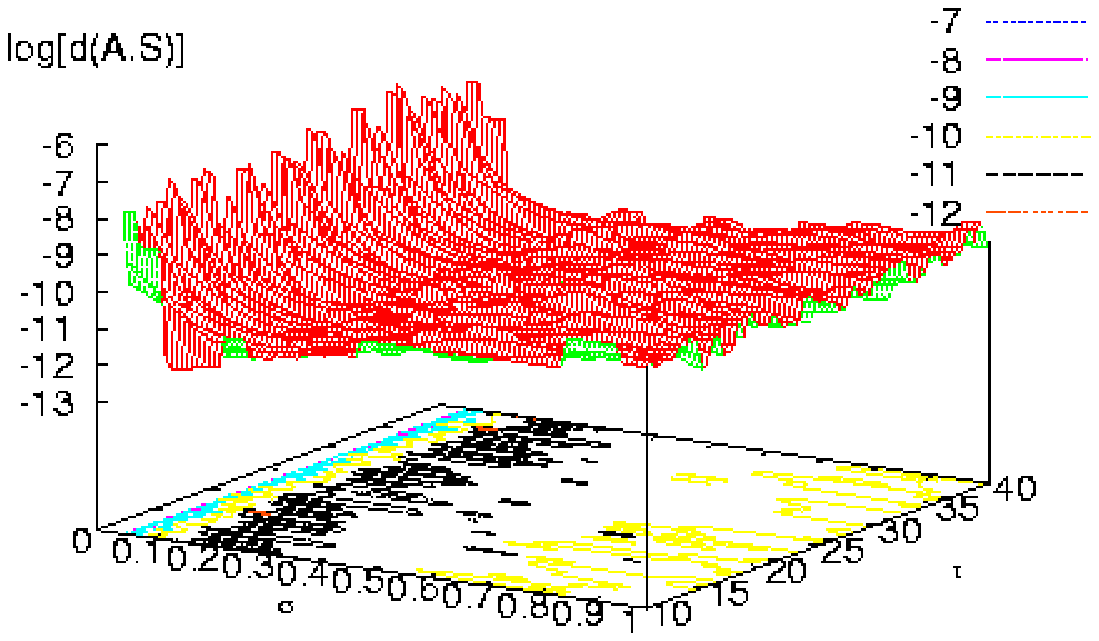}
\includegraphics[height=7cm,width=7cm,clip=false]{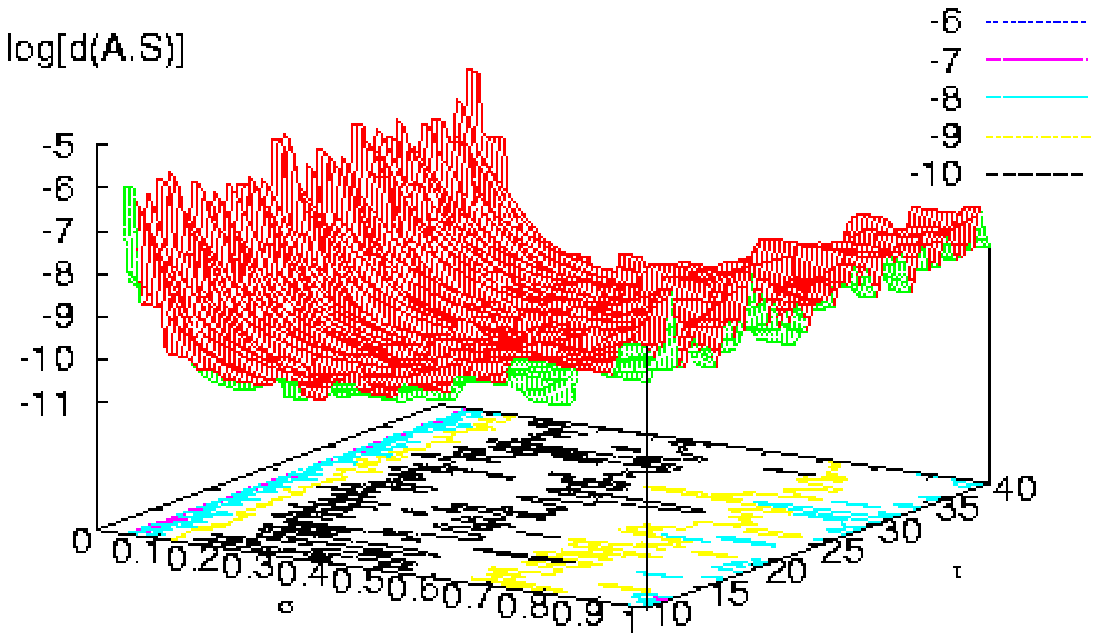}

\vspace{0.5cm}
\caption{Average value of $\log(\dAS)$  for the models I (top left)-II (top right)- III (bottom left)-
IV (bottom right) with $\sigma \in [0.01,1]$, $\tau_L \in [10,40] ms$. }
\label{Fdist}\end{center}
\end{figure}

\begin{figure}[ht!]
\begin{center}
\includegraphics[height=7cm,width=7cm,clip=false]{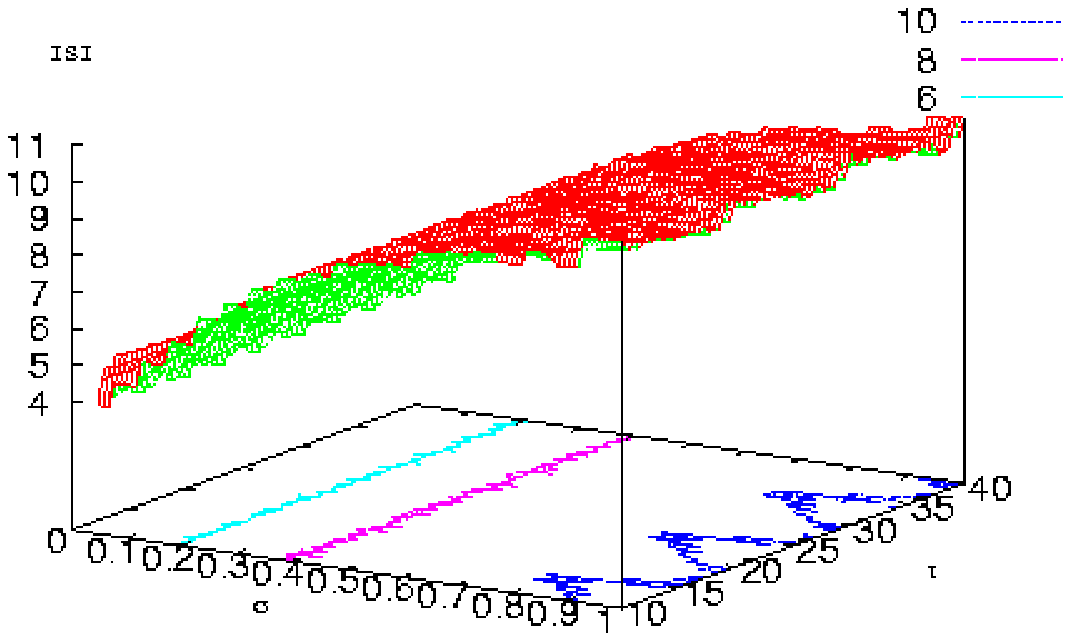}
\includegraphics[height=7cm,width=7cm,clip=false]{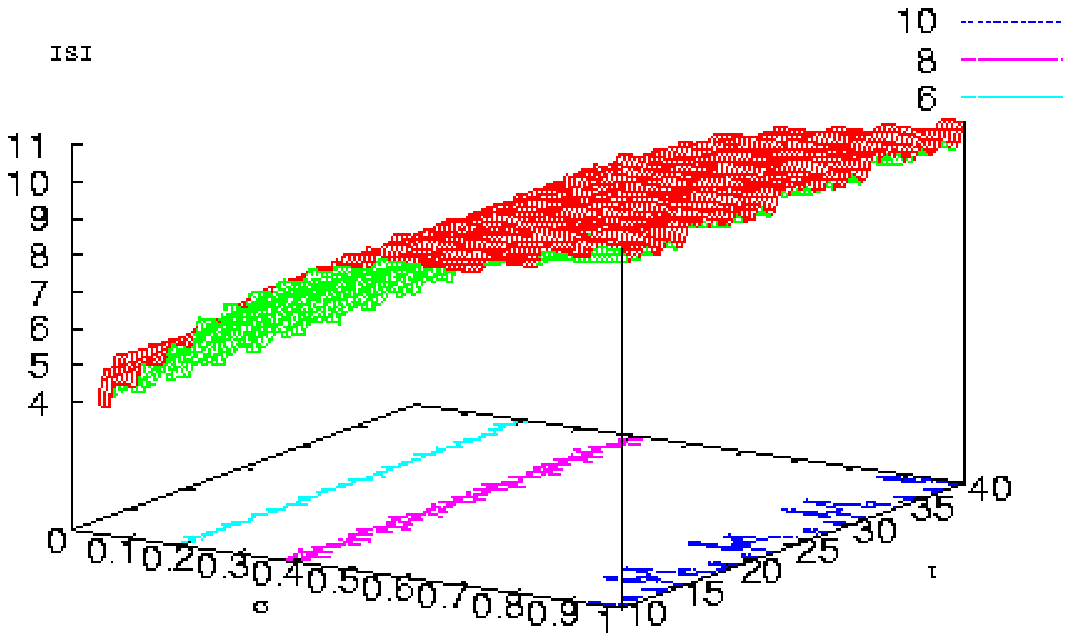}

\vspace{0.5cm}

\includegraphics[height=7cm,width=7cm,clip=false]{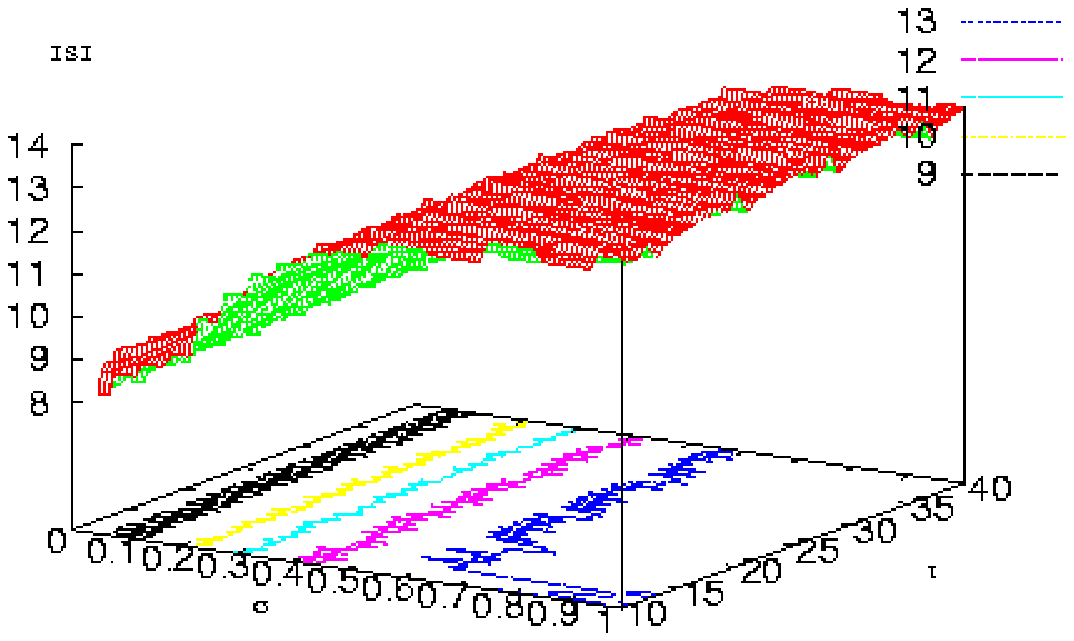}
\includegraphics[height=7cm,width=7cm,clip=false]{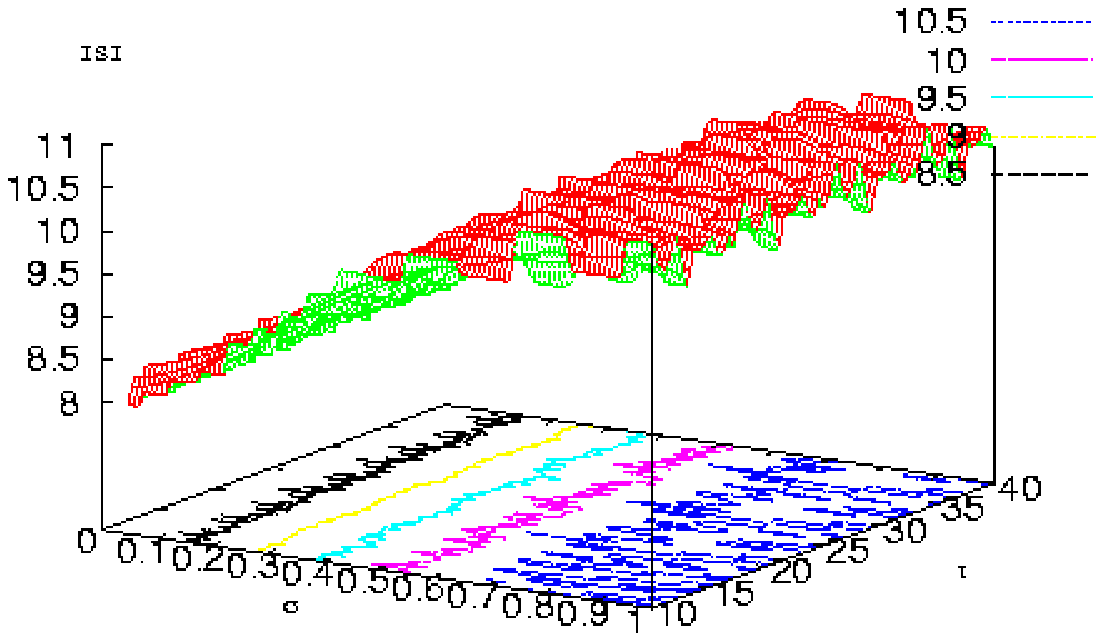}

\vspace{0.5cm}
\caption{Average value of the Inter Spike Interval  for the models I (top left)-II (top right)- III (bottom left)-
IV (bottom right), with $\sigma \in [0.01,1]$, $\tau_L \in [10,40] ms$. }
\label{FISI}\end{center}
\end{figure}

Then, we have considered the average value of the
current $\Jkto$, averaged over time, initial conditions and neurons and denoted by $I$ to
alleviate the notations (Fig. \ref{Fcurr}),
the logarithm of the distance $\dAS$ (Fig. \ref{Fdist}),
and the average Inter Spike Interval (ISI- Fig. \ref{FISI}),
for the four models.
The main observations are the following. Average current and Inter Spike Intervals
have essentially the same form for all models. This means that these quantities
are not really relevant if one wants to discriminate the various models
in their dynamical complexity. 
%
%%%%%%%%%%%%%%%%%%%% Cas 1cdf
%%%%%%%%%%%%%%%%%%%%%%%%%

%
%%%%%%%%%%%%%%%%%%%% Cas 1cdf
%%%%%%%%%%%%%%%%%%%%%%%%%

%
%%%%%%%%%%%%%%%%%%%% Cas 1cdf
%%%%%%%%%%%%%%%%%%%%%%%%%

%
%%%%%%%%%%%%%%%%%%%% Autres Cas
\begin{figure}[ht!]
\begin{center}
\includegraphics[height=7cm,width=7cm,clip=false]{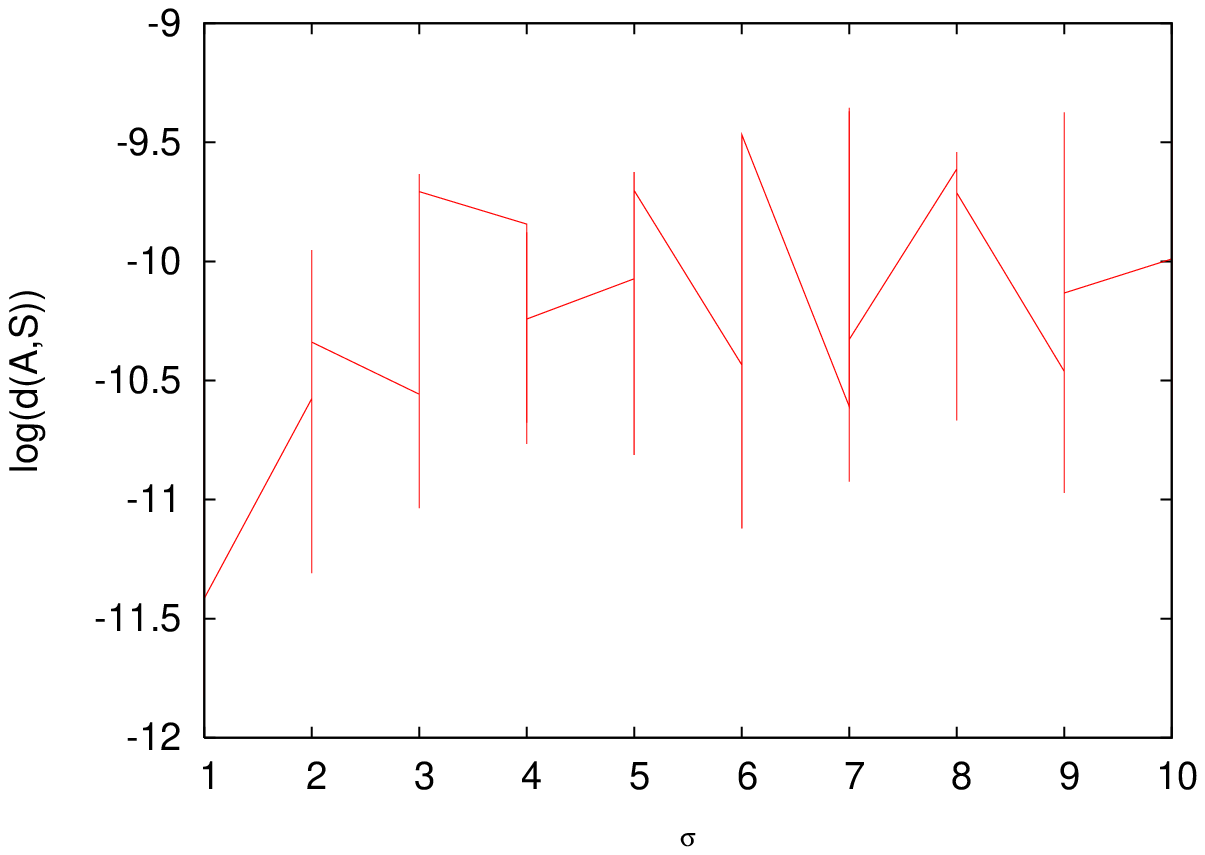}
\hspace{0.5cm}
\includegraphics[height=7cm,width=7cm,clip=false]{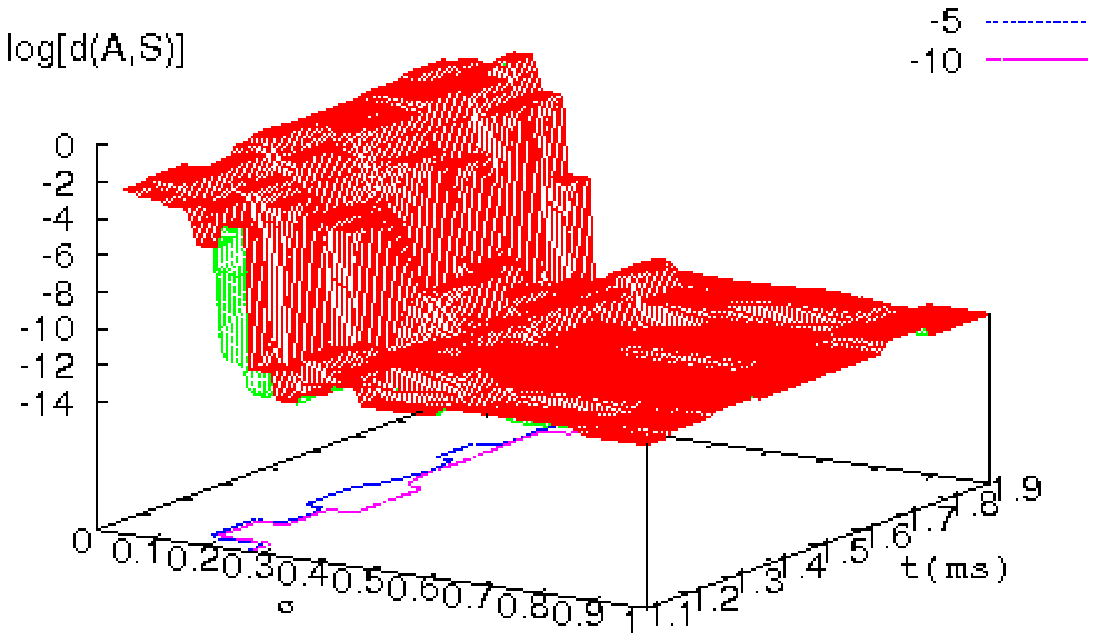}
\hspace{0.5cm}
\includegraphics[height=7cm,width=7cm,clip=false]{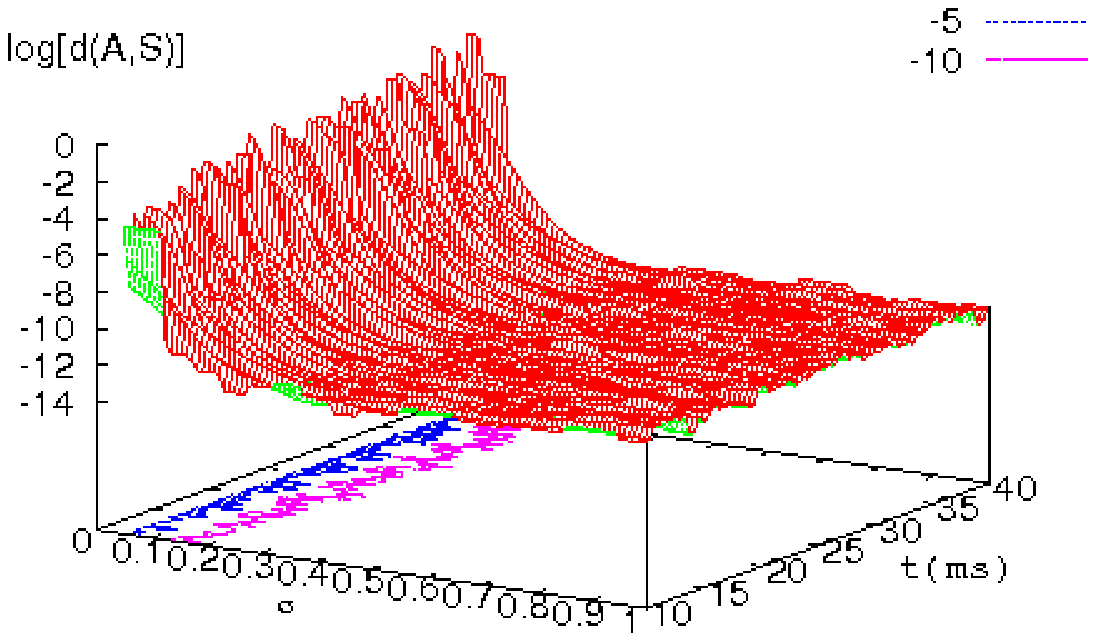}
\vspace{0.5cm}
\caption{Average value of $\log(\dAS)$  for the model I.
(left) $\sigma 
\in [1,10]$, $\tau \in [10,40]$ ms. We present here a projection
in the plane $\sigma,\log\dAS$, and the vertical bars correspond
to the  variations with $\tau_L$. It allows us to verify 
the stability of the previous result for higher variability  of the 
synaptic weights. (middle) $\tau_L \in [1,1 \dots 2] ms$
below the usual  $20 ms$ value, $\sigma \in [1,10]$.  Such range 
corresponds to cortical neurons in high-conductance state. It allows to 
check the behavior of $\dAS$ in this case.
(right) Sampling 
period of $1ms$, 
 in order to verify the robustness of the numerical results with respect 
to the sampling rate.}

\label{Fdistbis}\end{center}
\end{figure}
%%%%%%%%%%%%%%%%%%%%%%%%%

The observation of the distance $\dAS$ is quite more interesting.
First, in the four models, the distance becomes very small  when crossing
some ``critical region''  in the plane $\tau_L,\gamma$. This region has a regular
structure for the BMS model, but its structure seems more complex
for (\ref{DNN}). Note however that the numerical investigations used
here do not allow us to really conclude on this point.
The most remarkable fact is that, in models III and IV, the distance increases
when  $\sigma$  increases beyond this region, while
it does not in models I and II.
This corresponds to the following observation. When 
the $\dAS$ is small, one observes a complex dynamics with no apparent period.
One naturally concludes to a chaotic regime. As we saw, strictly speaking it is in fact periodic
but since periods are well beyond observable times, the situation is virtually
chaotic\footnote
{
Moreover, it is likely
that the phase space structure has some analogies with spin-glasses.
For example, if $\gamma=0$ the dynamics is essentially equivalent to the Kauffman's cellular
automaton \cite{kauffman:69}. It has been shown by Derrida and coworkers \cite{derrida-flyvbjerg:86,derrida-pomeau:86}
that the Kauffman's model has a structure similar to the Sherrington-Kirckpatrick spin-glass
model\cite{mezard:87}. The situation is even more complex when $\gamma \neq 0$.
It is likely that we have in fact a situation very similar to discrete time neural networks
with firing rates where a similar analogy has been exhibited \cite{cessac:94,cessac:95}.
}.
 When the distance increases, the orbits period decreases.
Therefore, there is a range of $\sigma$ values
where period become smaller than observational time and one concludes
that dynamics is periodic.

The situation is different for model I,II since the distance
does not apparently increases with $\sigma$. This suggests
that introducing conductance based synapses and currents enhances considerably
the width of the edge of chaos. On practical grounds, this means
that model I,II have the capacity to display a very large number
of distinct codes for wide choices of parameters.
This is somewhat expected since the opposite conclusion would mean
that introducing spike dependent conductances and current
does not increases the complexity and information capacity of the system.
But it is one thing to guess some behavior
and another thing to measure it. Our investigations on the
distance $\dAS$, a concept based on the previous mathematical
analysis, makes a step forward in this direction.

\begin{figure}[ht!]
\begin{center}
\begin{tabular}{cccc}
\parbox{0.5cm}{{\em I}~\\~\\~\\~\\~\\} &
\includegraphics[width=7cm,height=4cm,clip=false]{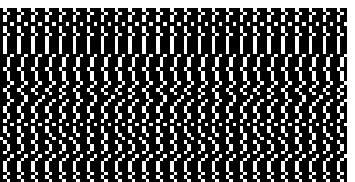} &
\includegraphics[width=7cm,height=4cm,clip=false]{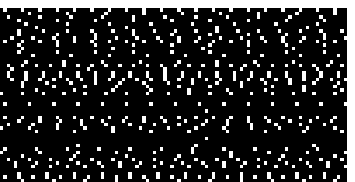} \\
\parbox{0.5cm}{{\em IV}~\\~\\~\\~\\~\\} &
\includegraphics[width=7cm,height=4cm,clip=false]{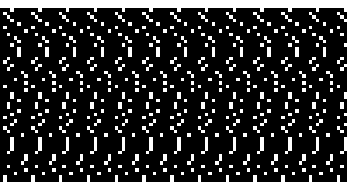} &
\includegraphics[width=7cm,height=4cm,clip=false]{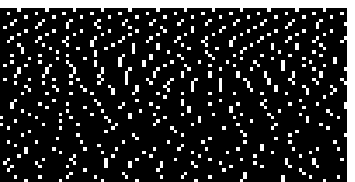} \\
& $\sigma=0.05$ & $\sigma=0.40$ \\
\end{tabular}

\vspace{0.5cm}
\caption{Examples of raster plots for the conductance based model ({\em Model I}, top row) and the leaky integrate and fire model ({\em Model IV}, bottom row).
A time window of $100$ samples is shown in each case. The control parameter is $\tau_L = 20ms$.
As visible in Fig.~\ref{Fdist}, $\sigma=0.05$ corresponds to a small order dynamics where the periodic behavior is clearly visible, and
$\sigma=0.40$ to the ``edge of chaos''. One blob width is 1msec}
\label{Fraster}\end{center}
\end{figure}

\begin{figure}[ht!]
\begin{center}
\begin{tabular}{cccc}
\parbox{0.5cm}{{\em I}~\\~\\~\\~\\~\\} &
\includegraphics[width=7cm,height=4cm,clip=false]{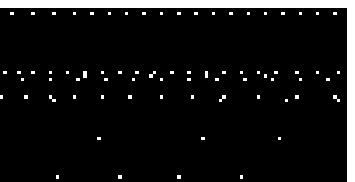} &
\includegraphics[width=7cm,height=4cm,clip=false]{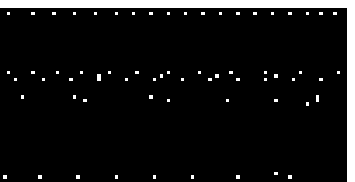} \\
\parbox{0.5cm}{{\em IV}~\\~\\~\\~\\~\\} &
\includegraphics[width=7cm,height=4cm,clip=false]{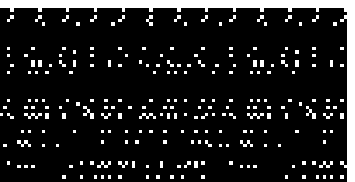} &
\includegraphics[width=7cm,height=4cm,clip=false]{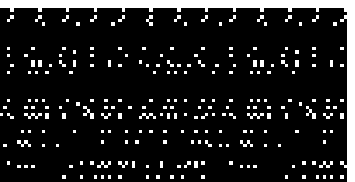} \\
& without noise & with noise \\
\end{tabular}

\vspace{0.5cm}
\caption{Raster plots for models I (upper row) and IV (lower row), with $\sigma=10.00$ and the same condition as in Fig. \ref{Fraster}.
First column: model I  and model  without noise. 
Second column: same realization of synaptic weights and same initial conditions but with a small amount of noise in the external current.
The noise is added to the membrane potential and its magnitude is very small ($10^{-4}\times \theta$). One blob width is 1msec.
}
\label{Fraster2}\end{center}
\end{figure}

One step further, we have represented examples
of raster plots  in Fig. \ref{Fraster} and \ref{Fraster2} for models I and IV.
The figure \ref{Fraster} essentially illustrates the discussion
above on the relation between the distance $\dAS$ and the dynamics;
for $\sigma=0.05$, where $\dAS$ is ``large'', and dynamics is periodic;
 and for $\sigma=0.4$, where $\dAS$ is small, and dynamics looks more chaotic,
for the two models. The difference between the two models becomes
more accurate as $\sigma$ increases. Fig. \ref{Fraster2}
represents  raster plots for models I,IV, with $\sigma=10$,
where we study the effect of a small amount of noise, of amplitude
$10^{-4} \times \theta$ in the external current. This has no
effect on model IV while it changes slightly the raster plot
for model I, as expected. There is another remarkable difference.
The code is sparser for model I than for model IV.
This suggests that model I is in some sense optimal with respect
to coding since it is able to detect very small changes in an input
but the changes is not drastic and the neural code remains very sparse.  

%%%%%%%%%%%%%%%%%%%%%%%%%

%%%%%%%%%%%%%%%%%%%%%%%%%

\su{Discussion} \label{y2disc}

We have thus
an operational definition for the ``edge of chaos''
where an ``order parameter'', the distance of orbits to 
the singularity has been defined. This parameter has a deep
meaning. It controls how much the system is sensitive to
perturbations. Such perturbations can be noise, but they
can also be a small variation in the external current,
corresponding e.g. to an input. If the amplitude
of this perturbation is smaller than $\dAS$ then 
it has no effect on the long term dynamics, and the neural
code (raster plot) is unchanged. On the other hand,
when the distance is small, even a tiny perturbation
 has a dramatic effect on the raster plot: the system
produces a different code. As a corollary, the effective
entropy is maximal when the distance is minimal. On practical
ground, having a positive distance with a large effective
entropy corresponds to situations where the system
is able to produce a large number of distinct codes within
the observational time, while this code is nevertheless
robust to small perturbations of the input. Thus, we have
a good compromise between the variability of the responses
to distinct inputs and robustness of the code when an input
is subject to small variations.\\

Several questions are now open. A first one concerns the way how we measured this distance.
We used a random sampling with \textit{independent} synaptic 
weights. But these weights are, in reality, highly correlated,
via synaptic plasticity mechanism. What is the effect
of e.g. STPD or Hebbian learning on the effective entropy is a perspective for a future
work.
Recent results in \cite{soula:05} and \cite{Siri:07a,Siri:07b}
suggest that synaptic plasticity reduces the entropy 
by diminishing the variability of raster plots and increasing
the robustness of the response to an input. Some general
(variational) mechanism could be at work here. This aspect
is under investigation.\\

Another important issue is the effect of noise. It is usual in neural network modeling to add Brownian noise to the
deterministic dynamics. This noise accounts for different effects
such as the diffusion of neurotransmitters involved in the synaptic
transmission, the degrees of freedom neglected by the model,
external perturbations, etc ... Though it is not evident that the ``real  noise''
is Brownian, using this kind of perturbations has the advantage
of providing a tractable model where  standard theorems in the theory
of stochastic processes \cite{touboul-faugeras:07} or methods in non equilibrium statistical physics (e.g. Fokker-Planck
equations \cite{brunel-hakim:99}) can be applied.

Though we do not treat explicitly this case in the present work, the formalism
has been designed to handle noise effects as well.
As a matter of fact, the effect of  Brownian noise on the dynamics of our model 
can be analyzed with standard techniques in probability theory
and stochastic perturbations of dynamical systems \cite{freidlin-wentzell:98}.
In particular, the probability distribution of the membrane potential trajectory
 can be obtained by using a discrete time version of Girsanov theorem \cite{samuelides-cessac:07}.
Noise  have several effects. Firstly,  the stochastic
trajectories stay around the unperturbed orbits until they jump to another
attraction basin, the characteristic time  depending on the noise intensity
(``Arrhenius law''). This has the effect of rendering the dynamics uniquely ergodic,
which somehow simplifies the statistical analysis.
The effect of noise will be essentially prominent in
 the region where $\dAS$ is small, leading to an effective initial
condition sensitivity and an effective positive Lyapunov exponent, that could
be computed using mean-field approaches \cite{cessac:95}.
It is possible to estimate the probability that a trajectory approaches
the singularity set $\cS$ within a finite time $T$ and a distance $d$ by using
Freidlin-Wentzell estimates \cite{freidlin-wentzell:98}. One can also construct a Markov chain for the transition
between the attraction basin of the periodic orbits of the unperturbed dynamics. 
The overall picture could be very similar (at least for BMS model)
to what happens when stochastically perturbing Kauffman's model \cite{golinelli-derrida:89},
with possibly a phase space structure reminiscent of spin-glasses (where noise
plays the role of the temperature).
This study is under investigations.\\

Yet another important issue relates to the fact that spikes can also be lost. This aspect is not yet taken into account in 
the present formalism, but annihilation of spikes is a future issue to address.

A final issue is the relation of this work with possible
biological observations. We would like in particular to come
back to the abstract notion of ghost orbit.
As said in the text, this notion corresponds to situation
where the membrane potential of some ``vicious'' neuron fluctuates below
the threshold, and approaches it arbitrary close, with no
possible anticipation of its first firing time. This leads to
an effective unpredictability in the network evolution,
since when this neuron eventually fire, it may drastically
change the dynamics of the other neurons, and therefore the observation
of the past evolution does not allow one to anticipate what will be
the future. In some sense, the system is in sort of a metastable state
but it is not in  a stationary state.
  
Now, the biological intuition tends to consider that a neuron cannot suddenly fire
after a very long time, unless its input changes. This suggests therefore that
``vicious'' neurons are biologically implausible. However, this argument,
to be correct, must precisely define what is a ``very long time''.
In fact, one has to compare the time scale of the experiment
to the characteristic time where the vicious neurons will eventually
fire. Note also that since only a very small portion of neurons can be observed e.g. in a given cortex area,
some ``vicious'' neurons could be present (without being observed since not firing),
with the important consequence discussed in this paper. The observation of ``temporarily silent'' neurons
which firing induces a large dynamic change would be an interesting issue in this context.

As a final remark we would like to point out the remarkable work of Latham and collaborators 
discussing the effects induced by the addition or removal of a single spike
in a raster plot. A central question is whether this ``perturbation'' (which is not necessarily ``weak'')
will have a dramatic effect on the further evolution (see \cite{latham-roth-etal:06} and
the talk of P. Latham available on line at $http://www.archive.org/details/Redwood\_Center\_2006\_09\_25\_Latham$).
Especially the questions
and discussions formulated during the talk of P. Latham are particularly salient in view of the
present work. As an additional remark note that
a perturbation may have an effect on trajectories but not on
the statistics build on these trajectories (e.g. frequency rates) \cite{cessac-sepulchre:06}.

\appendix

\su{Appendix.}

\ssu{Computation of $V_k(t+\delta)$}\label{SVtp}

Fix $t_1,t_2 \in [t,t+\delta[$. Set:

\beq
\nu_k(t_1,t_2,\tot)=e^{-\int_{t_1}^{t_2}g_k(s,\tos) \, ds}=e^{-\int_{t_1}^{t_2}g_k(s,\tot)ds},
\eeq

\nid where the last equality holds from our assumption that spikes
are taken into account at times multiples of $\delta$;
therefore $\tos=\tot, s \in [t,t+\delta[$.

We have:

 $$\nu_k(t_1,t_1,\tot)=1,$$

\beq\label{chainrule}
\nu_k(t_1,t_2,\tot)=\nu_k(t_1,t'_1,\tot)\nu_k(t'_1,t_2,\tot),
\eeq

\nid for $t'_1 \in [t,t+\delta[$.
Moreover:
 
$$\frac{d \nu_k(t_1,t_2,\tot)}{dt_1}=g_k(t_1,\tot)\nu_k(t_1,t_2,\tot),$$

This leads to:
 
$$\frac{d}{dt_1}\left(\nu_k(t_1,t_2,\tot) V_k(t_1)\right)=
\nu_k(t_1,t_2,\tot)
\left[\frac{dV_k}{dt_1}+ g_k(t_1,\tot) V_k(t_1)\right]
=\nu_k(t_1,t_2,\tot)i_k(t_1,\tot).$$

If neuron $k$ does not fire between $t$ and $t+\delta$ we have,
integrating the previous equation for  $t_1 \in [t,t+\delta[$
and setting $t_2=t+\delta$ :

\beq\label{DdiscdeltaND}
V_k(t+\delta)=\nu_k(t,t+\delta,\tot)V_k(t)+\int_{t}^{t+\delta} i_k(s,\tot) \nu_k(s,t+\delta,\tot) \, ds.
\eeq

\ssu{Proof of theorem \ref{Tomega}.}\label{Pomega}

The proof uses the following lemma.

\blem\label{Lomega}
Fix $\cF$ a subset of $ \left\{1 \dots N \right\}$ and let $\bcF$ be the complementary set of $\cF$. Call
$$\GFTe=
\left\{
\V \in \cM \left| 
\baR{ccc}
&(i) \ \forall i  \in \cF,&  \exists t \leq T, \mbox{such \ that} \ V_i(t) \geq \theta\\
&(ii) \ \forall j \in \bcF,& \exists t_0 \equiv t_0(\V,j) <  \infty ,  \mbox{such \ that} 
\ \forall t > t_0,  V_j(t) < \theta -\epsilon
\eaR
\right.
\right\}
$$
\nid then $\Omega(\GFTe)$, the $\omega$-limit set of $\GFTe$, is composed by finitely many
periodic orbits with a period $\leq T$. 
\elem

\bpr of theorem \ref{Tomega}

Note that there are finitely many subsets $\cF$ of $ \left\{1 \dots N \right\}$.
Note also that $\GFTe \subset \GFTpe$ and that $\GFTe \subset \GFTep$ whenever $\epsilon' < \epsilon$. 
We have therefore:

$$\cM \subset \bigcup_{\cF} \bigcup_{T>0} \bigcup_{\epsilon >0} \GFTe 
= \bigcup_{\cF}
 \Gamma_{\cF,+\infty,0}.$$

But, under hypothesis (1) and (2) of theorem \ref{Tomega}, there exists $\epsilon > 0, T <\infty $
such that $\cM=\bigcup_{\cF} \GFTe$ where the union on $\cF$ is finite.
Since $\F(\cM) \subset \bigcup_{\cF} \F(\GFTe)$, $\oM \subset  \bigcup_{\cF} \Omega(\GFTe)$.
Under lemma \ref{Lomega} 
$\oM$ is therefore a subset of a finite union of sets
containing  finitely many periodic orbits with a period $\leq T$.
\epr

\bpr of lemma \ref{Lomega}
Call $\PF$ (resp. $\PbF$) the projection onto the subspace generated by the basis vectors $\be_i, \ i \in \cF$
(resp.  $\be_j, \ j \in \bcF$) and set $\VF = \PF\V$ ($\VFb=\PbF \V$), $\FF=\PF\F$ ($\FFb=\PbF \F$).
Since each neuron $j \in \bcF$ is such that (\ref{Vkt}):

\beq\label{VjF}
V_j(t) =\sum_{n=1}^{t-1}  J_k(n,\ton)\prod_{s=n+1}^{t-1} \gks(1-\omega_k(s))
< \theta-\epsilon,
\eeq
for $t$ sufficiently large, (larger than the last (finite) firing time $t_j$),
  these neurons do not act on the other neurons 
and their membrane potential
is only a function of the synaptic current generated by the neurons $\in \cF$. 
Thus, the asymptotic dynamics 
is generated by the neurons $i \in \cF$.  Then,
$\forall \V \in \oGFTe$, $\VF(t+1)=\FF[\VF(t)]$ and
$\VFb(t+1)=\FFb[\VF(t)]$. One can therefore focus the analysis of
the $\omega$ limit set to its projection $\oFTe=\PF\oGFTe$ (and infer the dynamics
of the neurons $j \in \bcF$ via (\ref{VjF})). 

Construct now the partition $\cPT$, with convex elements
given by $\MoT$, where $T$ is the same as
in the definition of $\GFTe$. By construction, $\FTp$ is continuous on each element $\cPT$
and fixing $\MoT$ amounts to fix the affinity constant of $\FTp$. 
By definition of $T$, $\DFFTpV$,    the derivative of $\FTpF$ at $\V$, 
has all its eigenvalues equal to $0$ whenever $\V \in \oFTe$ (prop. \ref{PContract}). 
Therefore $\FTpF[\MoT \cap \oFTe]$
is a point. Since 
$$\FTpF(\cM \cap \oFTe)=\FTpF\left(\bigcup \MoT \cap \oFTe \right) \subset 
\bigcup \FTpF\left( \MoT \cap \oFTe \right),$$
\nid  the image of $\oFTe$ under 
$\FTpF$ is a finite union of points belonging 
to $\cM$. Since, 
$\oFTe$ is invariant, this a finite union of
points, and thus  a finite union of periodic
orbits.

 The dynamics of neurons
$\in \bcF$ is driven by the periodic dynamics of firing neurons
and it is easy to see that their trajectory is asymptotically periodic. 
Finally, since $\cM=\cup_{\cF} \GFTe$ 
the $\omega$ limit set of $\cM$
is a finite union of periodic orbits.  \epr

\ssu{Average of a function.}

Since the dynamics is not uniquely ergodic (there are typically many
periodic attractors), one has to be
careful with the notion of average of a function $\phi$. 
We have first to perform a time average
for each attractor $i$,
$\bar{\phi}^{(i)} = \lim_{T  \to \infty} \sum_{t=1}^T \phi(\V^{(i)}(t))$,
where $\V^{(i)}$ is an initial condition in
the attraction basin of attractor $i$.
Then, we have to average over all attractors, with a weight
corresponding to the Lebesgue measure $\mu^{(i)}$ of its attraction basin.
This gives:

\beq\label{Average}
\left<\phi \right> = 
\frac{1}{\cN}\sum_{i=1}^\cN \mu^{(i)}\bar{\phi}^{(i)}
\eeq

\nid where $\cN$ is the number of attractors.

\bibliographystyle{apacite}
\bibliography{biblio}

%{\scriptsize  \bibliographystyle{apacite} \bibliography{odyssee,biblio}}
\end{document}